\documentclass[amsmath,amssymb,prb,superscriptaddress,preprint,longbibliography]{revtex4-1}

\usepackage{amsmath,amsfonts,amssymb}
\usepackage{graphicx}
\usepackage{hyperref}
\hypersetup{colorlinks=true, citecolor=blue, urlcolor=blue, linkcolor=blue}
\usepackage[usenames,dvipsnames]{xcolor}
\usepackage{soul}
\usepackage{braket}
\usepackage{bm}
\usepackage{upgreek}
\usepackage{siunitx}

\newcommand{\MC}[1]{\mathcal{#1}}
\newcommand{\MR}[1]{\mathrm{#1}}
\newcommand{\ud}{\mathrm{d}}
\newcommand{\iu}{\mathrm{i}}
\newcommand{\Tr}{\mathrm{Tr}}

\newcommand{\VEC}[1]{\mathbf{#1}}

\newcommand{\NORM}[1]{\lvert #1 \rvert}
\newcommand{\paulivec}{\bm{\upsigma}}

\begin{document}

\title{Insights into the orbital magnetism of noncollinear magnetic systems}

\author{Manuel dos Santos Dias}\email{m.dos.santos.dias@fz-juelich.de}
\author{Samir Lounis}\email{s.lounis@fz-juelich.de}
\affiliation{Peter Gr\"{u}nberg Institut and Institute for Advanced Simulation, Forschungszentrum J\"{u}lich \& JARA, D-52425 J\"{u}lich, Germany}

\date{\today}

\begin{abstract}
The orbital magnetic moment is usually associated with the relativistic spin-orbit interaction, but recently it has been shown that noncollinear magnetic structures can also be its driving force.
This is important not only for magnetic skyrmions, but also for other noncollinear structures, either bulk-like or at the nanoscale, with consequences regarding their experimental detection.
In this work we present a minimal model that contains the effects of both the relativistic spin-orbit interaction and of magnetic noncollinearity on the orbital magnetism.
A hierarchy of models is discussed in a step-by-step fashion, highlighting the role of time-reversal symmetry breaking for translational and spin and orbital angular motions.
Couplings of spin-orbit and orbit-orbit type are identified as arising from the magnetic noncollinearity.
We recover the atomic contribution to the orbital magnetic moment, and a nonlocal one due to the presence of circulating bound currents, exploring different balances between the kinetic energy, the spin exchange interaction, and the relativistic spin-orbit interaction.
The connection to the scalar spin chirality is examined.
The orbital magnetism driven by magnetic noncollinearity is mostly unexplored, and the presented model contributes to laying its groundwork.
\end{abstract}

\maketitle

\section{Introduction}\label{sec:intro}
Magnetic skyrmions\cite{Bogdanov1989} are a kind of topological twist in a ferromagnetic structure, with unusual properties.\cite{Nagaosa2013}
They have been found in bulk samples and in thin films, also at room temperature\cite{Muhlbauer2009,Yu2010a,Heinze2011,Shibata2013,Boulle2016,Moreau-Luchaire2016,Woo2016}.
When electrons travel throught the noncollinear magnetic structure of the skyrmion, they experience emergent electromagnetic fields.\cite{Xiao2010a,Franz2014,Sitte2014}
This strong coupling between the electronic and magnetic degrees of freedom leads to very efficient motion of skyrmions with electric currents.~\cite{Jonietz2010}
It also generates a topological contribution to the Hall effect, a transport signature of a skyrmion-hosting sample,\cite{Neubauer2009,Schulz2012} and was shown to enable the electrical detection of an isolated skyrmion.~\cite{Crum2015,Hanneken2015}
The link between the magnetic structure and orbital electronic properties was explored for other kinds of magnetic systems before,\cite{Shindou2001,Tatara2002,Tatara2003,Tatara2003a,Nakamura2003a,Bulaevskii2008} with renewed interest since the experimental discovery of skyrmions.~\cite{Hamamoto2015,Hoffmann2015,Yin2015,Dias2016,Hanke2016,Goebel2017,Lux2017}

Inspired by the investigations of nanosized skyrmions in the PdFe/Ir(111) system,~\cite{Romming2013,Romming2015} we uncovered another manifestation of their topological nature: a new kind of orbital magnetism.
In Ref.~\onlinecite{Dias2016}, magnetic trimers and skyrmion lattices were compared, and the orbital magnetic moment was shown to have two contributions: a spin-orbit-driven one and a scalar-chirality-driven one.
The minimum number of magnetic atoms needed for a non-vanishing scalar chirality is three, $\VEC{n}_1 \cdot (\VEC{n}_2 \times \VEC{n}_3) \neq 0$, with $\VEC{n}_i$ the orientation of their respective spin magnetic moments.
This means that the magnetic structure is noncoplanar, a requirement for the appearance of this new kind of orbital magnetism.
Magnetic trimers were analyzed in detail within density functional theory (DFT), before considering a skyrmion lattice meant to mimic PdFe/Ir(111).
Although some calculations were feasible with DFT, to address larger skyrmion sizes a minimal tight-binding model was constructed from the DFT data.
This model reproduced both contributions to the local orbital moment, and showed that the sum of all scalar-chirality-driven contributions leads to a topological orbital magnetic moment for a skyrmion lattice.
The goal of the present paper is to get more insight into the physical mechanisms driving the orbital magnetism of systems in which both the relativistic spin-orbit interaction (RSOI) and a noncollinear magnetic structure coexist.
To this end, we reprise our tight-binding model of Ref.~\onlinecite{Dias2016} but now applied to magnetic trimers, and present a walkthrough of the different sources of orbital magnetism in this model.

In classical physics, the orbital magnetic moment arises from the presence of bound currents in a given material.
This is a consequence of the continuity equation for the electronic charge density in equilibrium:
\begin{equation}
  0 = \frac{\partial\rho}{\partial t} = -\nabla\cdot\VEC{j} \;\;\Longrightarrow\;\; \VEC{j}(\VEC{r}) = \nabla \times \VEC{m}(\VEC{r}) \quad.
\end{equation}
From the quantum-mechanical point of view, if the ground state supports such a finite bound current, time-reversal symmetry must be broken.
Magnetic materials naturally break time-reversal symmetry, due to the existence of ordered spin magnetic moments stabilized by exchange interactions.
As long as correlation effects are only moderately important, the orbital magnetic moment is usually assumed to be due to the RSOI, which can be introduced either in atomic or in Rashba-like form,~\cite{Manchon2015}
\begin{equation}
  \MC{H}_{\MR{SOI}} \propto \VEC{L} \cdot \paulivec \quad\MR{or}\quad (\VEC{E} \times \VEC{p}) \cdot \paulivec \quad.
\end{equation}
Here $\VEC{L} = \VEC{r} \times \VEC{p}$ is the atomic orbital angular momentum operator, $\paulivec$ the vector of Pauli matrices, $\VEC{E}$ the electric field and $\VEC{p} = -\iu\,\hbar\,\bm{\nabla}$ the linear momentum operator.
For an atom both forms are equivalent.
It is the coupling between the finite spin moment in a magnetic material and the orbital degrees of freedom depending on $\VEC{p}$ or $\VEC{L}$ that leads to the finite orbital moment.~\cite{Skomski2008}

Recently another way of coupling spin and orbital degrees of freedom has been identified and explored.~\cite{Xiao2010a,Sitte2014}
Consider the non-interacting electron hamiltonian consisting of a kinetic term and a spin exchange coupling to an underlying magnetic structure:
\begin{equation}\label{eq:freemodel}
  \MC{H} = \frac{\VEC{p}^2}{2m}\,\sigma_0 + J\,\VEC{m}(\VEC{r}) \cdot \paulivec \quad.
\end{equation}
Here $\sigma_0$ is the unit spin matrix and $J$ is the strength of the exchange coupling.
The magnetization field is $\VEC{m}(\VEC{r}) = m(\VEC{r})\,\VEC{n}(\VEC{r})$, with its spatially varying magnitude $m(\VEC{r})$ and direction $\VEC{n}(\VEC{r})$.
For collinear magnetic systems, such as ferromagnets or simple antiferromagnets, $\VEC{m}(\VEC{r}) = m(\VEC{r})\,\VEC{n}_z$, where $\VEC{n}_z$ is the direction of the ferromagnetic or staggered magnetization, respectively (accordingly, we allow $m(\VEC{r})$ to be negative).
Then the eigenstates of the system can be labelled with the usual `up' and `down' eigenspinors of $\sigma_z$, and we get two decoupled hamiltonians, one for spin-up and another for spin-down.
In this way the magnetic order decouples from the orbital degrees of freedom, if the RSOI is not considered.

If a common spin quantization axis cannot be chosen, i.e.~the system has a noncollinear magnetic structure, such a coupling is indeed present.
To see how this arises, consider the unitary transformation that at every point in space diagonalizes the exchange term,
\begin{equation}
  U^\dagger(\VEC{r})\,\VEC{m}(\VEC{r}) \cdot \paulivec\;U(\VEC{r}) = m(\VEC{r})\,\sigma_z
  \;\;\Longrightarrow\;\; U(\VEC{r}) = e^{\iu\VEC{w}(\VEC{r}) \cdot \paulivec} \quad .
\end{equation}
The vector $\VEC{w}(\VEC{r})$ describes the spin rotation in the axis-angle representation, but its explicit form is not required for the present argument, only the fact that it must have a spatial dependence for a noncollinear magnetic structure.
If this unitary transformation is applied to the whole hamiltonian, it effects a SU(2) gauge transformation~\cite{Tokatly2008,Berche2012,Berche2013} with the result
\begin{equation}
  \MC{H}' = U^\dagger \MC{H}\,U = \frac{\boldsymbol{\Pi}^2}{2m} + J\,m(\VEC{r})\,\sigma_z \quad,\qquad
  \Pi_{\mu} = p_\mu\,\sigma_0 + \sum_\nu\left(\hbar\,\partial_\mu w_\nu\right)\sigma_\nu
  = p_\mu\,\sigma_0 + \sum_\nu A_{\mu\nu}\,\sigma_\nu \quad .
\end{equation}
The kinetic momentum $\Pi_\mu$ consists of the canonical momentum $p_\mu$ and a vector potential $A_{\mu\nu}$ that couples to the spin, with $\mu,\nu=x,y,z$.
The kinetic energy now has four contributions:
\begin{equation}
  \sum_\mu\Pi_\mu^2 = \sum_\mu p_\mu^2\,\sigma_0 + 2\sum_{\mu\nu} A_{\mu\nu}(\VEC{r})\,p_\mu\,\sigma_\nu
  - \iu\,\hbar\sum_{\mu\nu}\left(\partial_\mu A_{\mu\nu}(\VEC{r})\right)\sigma_\nu
  + \sum_{\mu\nu} \big(A_{\mu\nu}(\VEC{r})\big)^2 \sigma_0 \quad .
\end{equation}
The first term is the usual spin-independent contribution, the second term is a spin-orbit interaction (coupling spin to linear momentum), the third is a Zeeman-like contribution, and the fourth is a spin-independent potential-like contribution.
We thus see that noncollinear magnetic structures lead to emergent fields that couple the spin and orbital degrees of freedom.
This line of reasoning has been very successful in explaining the emergent electrodynamics of slowly-varying magnetic textures.~\cite{Nagaosa2013,Sitte2014}

When both kinds of spin-orbit interaction are at play, the one of relativistic origin and the one arising from a noncollinear magnetic structure, they compete with each other, and a unified picture can only be given for special limiting cases.
First-principles electronic structure calculations provide both qualitative and quantitative insights.
Here we adopt the tight-binding model of Ref.~\onlinecite{Dias2016} to analyze the smallest system for which both contributions to the orbital moment are present: a magnetic trimer.

This paper is organized as follows.
Section~\ref{sec:tbmodel} introduces the model and its ingredients, and then a step-by-step construction of the eigenstates with broken time-reversal symmetry is provided.
First, in Sec.~\ref{sec:model1} we explore how a magnetic field breaks the translational symmetry of the trimer.
Then we show in Sec.~\ref{sec:model2} how a noncollinear magnetic structure produces orbital effects analogous to those of an external magnetic field.
Finally, we combine noncollinear magnetism and the atomic spin-orbit interaction in Sec.~\ref{sec:model3}, drawing parallels between the relativistic and the noncollinear sources of orbital magnetism.
We discuss our results and present our conclusions in Sec.~\ref{sec:conclusions}.

\section{Tight-binding model for noncollinear magnetic structures}\label{sec:tbmodel}
In Ref.~\onlinecite{Dias2016} the following minimal tight-binding model was introducted, to describe two magnetic $d$-bands experiencing the effects of the relativistic spin-orbit interaction and of a noncollinear magnetic structure:
\begin{equation}\label{eq:tbmodel}
  \MC{H} = \MC{H}_{\MR{kin}} + \MC{H}_{\MR{mag}} + \MC{H}_{\MR{soi}} \quad.
\end{equation}
The kinetic energy is given by 
\begin{equation}
  \MC{H}_{\MR{kin}} = \sum_{i,j \neq i}\sum_{mm's} c_{ims}^\dagger\,t_{im,jm'}\,c_{jm's} \quad.
\end{equation}
Here $c_{ims}^\dagger$ creates an electron on atomic site $i$ and on the $d$-orbital labelled $m$, with spin projection $s$.
The detailed form of the hopping matrices $t_{im,jm'}$ is presented later.

The coupling to a background magnetic structure is described by
\begin{equation}
  \MC{H}_{\MR{mag}} = J\sum_i\sum_{mss'}c_{ims}^\dagger\,\VEC{n}_i \cdot \paulivec_{ss'}\,c_{ims'} \quad,
\end{equation}
with $J$ the strength of the coupling, and $\VEC{n}_i$ the unit vector describing the direction of the background magnetic structure on every atomic site.
We see that $\MC{H}_{\MR{kin}} + \MC{H}_{\MR{mag}}$ is the tight-binding equivalent of the model of Eq.~\ref{eq:freemodel} that was discussed in the introduction.

The RSOI is considered in atomic form,
\begin{equation}
  \MC{H}_{\MR{soi}} = \xi\sum_i\sum_{mss'}c_{ims}^\dagger\,\VEC{L}_{mm'} \cdot \paulivec_{ss'}\,c_{ims'} \quad,
\end{equation}
with $\xi$ its coupling strength and $\VEC{L}_{mm'}$ the matrix elements of the atomic orbital angular momentum operator for the two $d$-orbitals in the model.

Time reversal symmetry is usually described by the antiunitary operator $\MC{T} = \iu\,\sigma_y\,\MC{K}$.~\cite{Sakurai1994}
Here $\MC{K}$ takes the complex conjugate of the spinor wavefunction it is applied to, while $\iu\,\sigma_y$ ensures that the spin is also reversed.
The hamiltonian is time-reversal invariant if it commutes with it, $\left[\MC{T},\MC{H}\right] = 0$.
The action of $\MC{T}$ is illustrated in the following example:
\begin{equation}
  \Psi_{\VEC{k}\uparrow}(\VEC{r}) = e^{\iu\VEC{k}\cdot\VEC{r}} \begin{pmatrix} 1 \\ 0 \end{pmatrix} \;\;\Longrightarrow\;\;
  \MC{T}\,\Psi_{\VEC{k}\uparrow}(\VEC{r}) = e^{-\iu\VEC{k}\cdot\VEC{r}} \begin{pmatrix} 0 & 1 \\ -1 & 0 \end{pmatrix} \begin{pmatrix} 1 \\ 0 \end{pmatrix}
  = -e^{-\iu\VEC{k}\cdot\VEC{r}} \begin{pmatrix} 0 \\ 1 \end{pmatrix}
  = -\Psi_{-\VEC{k}\downarrow}(\VEC{r}) \quad .
\end{equation}
It is more helpful to think of $\MC{T}$ as reversing the state of motion.
We could then write this operator as $\MC{T} = \MC{T}_{\MR{P}}\,\MC{T}_{\MR{S}}$, where $\MC{T}_{\MR{P}}$ reverses the orbital part of the motion ($\VEC{k} \Rightarrow -\VEC{k}$ in the example) and $\MC{T}_{\MR{S}}$ reverses the spin angular momentum ($\uparrow\;\Rightarrow\;\downarrow$ in the example).~\cite{Haake2010}
Hamiltonians describing magnetic systems are typically not time-reversal invariant, either due to the presence of external magnetic fields or to the exchange interactions that stabilize the magnetic ground state.
Whether they might still be invariant under reversal of the orbital motion, $\MC{T}_{\MR{P}}$, is only clear in the momentum representation, as it amounts to $\MC{H}(-\VEC{p},\paulivec) = \MC{H}(\VEC{p},\paulivec)$.
An alternative is to choose basis functions that are not invariant under either $\MC{T}_{\MR{P}}$ or $\MC{T}_{\MR{S}}$, as shown by the combination of a plane-wave with a spinor in the example.~\cite{Haake2010}
This is the strategy that will be employed in the following.

\section{The trimer with simple hopping}\label{sec:model1}
\begin{figure}[b]
  \centering
  \includegraphics[width=\textwidth]{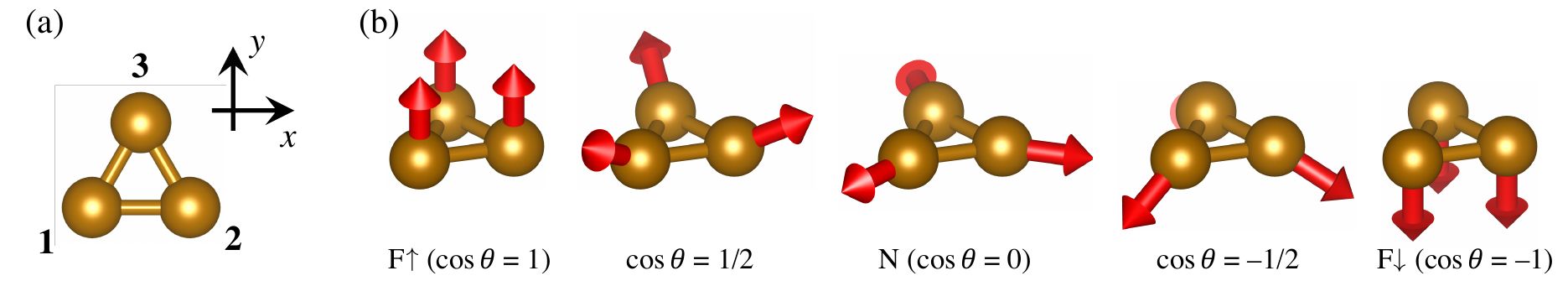}
  \caption{\label{fig:trimer} The magnetic trimer.
  (a) Atomic structure and choice of coordinate axes, with golden spheres representing the atomic sites.
  (b) Some magnetic structures described by Eq.~\eqref{eq:magdir}, with the choice of angles discussed in the text.
  The red arrows the local orientation of the magnetic structure.
  The structures with $\cos\theta = \pm 1$ are ferromagnetic, $\cos\theta = 0$ is the antiferromagnetic N\'eel structure, and the others are noncollinear structures.}
\end{figure}

Consider three identical atomic sites forming an equilateral triangle with sides taken as the unit of length, $a = 1$, as shown in Fig.~\ref{fig:trimer}(a).
To uncover the role of the electronic motion around the trimer, we first consider a simplified model with one orbital per site and no spin dependence.
Let $\ket{i}$ be a basis state for one electron being on atom $i$.
In this basis, the hamiltonian is given by
\begin{equation}\label{eq:model1}
  \renewcommand{\arraystretch}{1.2}
  \begin{array}{r|lll}
    \MC{H}_{\MR{kin}} & \ket{1} & \ket{2} & \ket{3} \\
    \hline
    \bra{1} & 0 & \bar{t} & t \\
    \bra{2} & t & 0 & \bar{t} \\
    \bra{3} & \bar{t} & t & 0
  \end{array}
  \quad\Longrightarrow\quad \MC{H}_{\MR{kin}} = t\,\MC{R}_0 + \bar{t}\,\MC{R}_0^\dagger \quad .
\end{equation}
$t = \NORM{t}\,e^{\iu\alpha}$ is the hopping amplitude for counterclockwise hops around the triangle, and its complex conjugate $\bar{t}$ is the one for clockwise hops.
The complex hopping breaks the symmetry of translational motion, and could be due to the presence of a magnetic flux threading the triangle.

The operator $\MC{R}_0$ generates the counterclockwise hops,
\begin{equation}
  \MC{R}_0 = \begin{pmatrix} 0 & 0 & 1 \\ 1 & 0 & 0 \\ 0 & 1 & 0 \end{pmatrix} \quad,\qquad \MC{R}_0 \ket{i} = \ket{i{+}1}
  \quad,\qquad \MC{R}_0^\dagger = \MC{R}_0^{-1} \quad,\qquad \MC{R}_0^\dagger \ket{i} = \ket{i{-}1} \quad ,
\end{equation}
and its spectral representation is
\begin{equation}\label{eq:eigenstates_model1}
  \MC{R}_0 = \sum_k e^{-\iu\frac{2\pi k}{3}} \ket{k}\!\bra{k} \quad,\qquad
  \ket{k} = \frac{1}{\sqrt{3}}\left(\ket{1} + e^{\iu\frac{2\pi k}{3}} \ket{2} + e^{-\iu\frac{2\pi k}{3}} \ket{3}\right) \quad,\qquad k \in \{0,\pm1\} \quad.
\end{equation}
Note that these basis states are not invariant under reversal of the translational motion, $\MC{T} \ket{k} = \MC{T}_{\MR{P}} \ket{k} = \ket{-k}$, as the clockwise motion is the time-reversed form of the counterclockwise motion.
The state $k=0$ corresponds to no overall translational motion, so it equals itself under $\MC{T}_{\MR{P}}$.

The hamiltonian commutes with $\MC{R}_0$, so from $\MC{H}_{\MR{kin}}\ket{k} = E_k\ket{k}$ we find the eigenenergies
\begin{equation}\label{eq:energies_model1}
  E_k = 2\,\NORM{t}\cos\left(\frac{2\pi k}{3}-\alpha\right) \quad.
\end{equation}
As this model is equivalent to a linear chain of three atoms with periodic boundary conditions, we shall call $k$ the ring momentum, which characterizes the translational motion of each eigenstate.
We can also define momentum raising and lowering operators, which will be very useful in the following sections:
\begin{equation}\label{eq:kpm}
  \MC{R}_{\pm} = \begin{pmatrix} 1 & 0 & 0 \\ 0 & e^{\pm\iu\frac{2\pi}{3}} & 0 \\ 0 & 0 & e^{\mp\iu\frac{2\pi}{3}}\end{pmatrix}
  \quad,\qquad \MC{R}_{\pm}\ket{k} = \ket{k{\pm}1} \quad,\qquad (\MC{R}_{\pm})^3\ket{k} = \ket{k} \quad .
\end{equation}
The last equality is due to the periodicity of the phase, $k \pm 3 = k$.

A uniform magnetic field perpendicular to the plane of the trimer is a simple example of broken time-reversal symmetry.
In the symmetric gauge, with the origin at the center of the triangle, the vector potential is given by
\begin{equation}
  \VEC{A}(\VEC{r}) = -\frac{B}{2}\,\VEC{r} \times \VEC{n}_z \quad.
\end{equation}
The Peierls substitution~\cite{Peierls1933,Boykin2001,Azpiroz2014} provides the phase acquired by an electron hopping from site $j$ to site $i$:
\begin{equation}\label{eq:tbpeierls}
  \alpha_{ij} = \frac{2\pi}{\Phi_0}\int_{\VEC{r}_j}^{\VEC{r}_i}\!\!\!\!\ud\VEC{r}\cdot\VEC{A}(\VEC{r})
  = \chi_{ij}\,\frac{2\pi}{\Phi_0}\,\frac{S}{3}\,B = \chi_{ij}\,\frac{2\pi}{3}\,\frac{\Phi_{\MR{B}}}{\Phi_0} = \chi_{ij}\,\alpha
  \;\;\Longrightarrow\;\; t_{ij} = \NORM{t}\,e^{\iu \chi_{ij} \alpha} \quad,
\end{equation}
with the path integral evaluated on the straight line connecting the sites.
$S = \sqrt{3}\,a^2/4 \sim \SI{0.1}{\nano\meter^2}$ is the area of the triangle, for typical bond lengths.
The sign of the result is $\chi_{ij} = 1$ if $i$ is a neighbor of $j$ in the counterclockwise sense, and $\chi_{ij} = -1$ for the clockwise sense.
$\Phi_0 = h/e \approx 4 \times 10^3\,\si{\tesla\,\nano\meter^2}$ is the magnetic flux quantum, and $\Phi_{\MR{B}}$ the actual magnetic flux threading the triangle.
Due to the magnetic field, hopping in a clockwise sense is no longer equivalent to hopping in a counterclockwise sense, and this leads to $E_{-k} \neq E_k$, as already derived above.

As $\alpha$ is proportional to the magnetic field $B$, we define the orbital magnetic moment operator as
\begin{equation}
  \MC{M} = -\frac{\partial\MC{H}}{\partial \alpha}
  = -\iu\,\NORM{t} \left(e^{\iu\alpha}\,\MC{R}_0 - e^{-\iu\alpha}\,\MC{R}_0^\dagger\right) \quad .
\end{equation}
It represents the net current flowing around the triangle, and can be evaluated for each eigenstate using the eigenfunctions given in Eq.~\eqref{eq:eigenstates_model1},
\begin{equation}\label{eq:magmoms_model1}
  M_k = -2\,\NORM{t} \sin\left(\frac{2\pi k}{3}-\alpha\right) = -\frac{\partial E_k}{\partial \alpha} \quad .
\end{equation}
The last equality is a nice illustration of the Hellmann-Feynman theorem.~\cite{Hellmann1937,Feynman1939}
The following properties will be useful to simplify certain matrix elements appearing in the next sections:
\begin{subequations}\label{eq:ekprops}
\begin{equation}
  E_{k{-}1} + E_{k} = -E_{k{+}1} \quad,\qquad M_{k{-}1} + M_{k} = -M_{k{+}1} \quad,
\end{equation}
\begin{equation}
  E_{k{-}1} - E_{k} = \sqrt{3}\,M_{k{+}1} \quad,\qquad M_{k{-}1} - M_{k} = -\sqrt{3}\,E_{k{+}1} \quad .
\end{equation}
\end{subequations}

\begin{figure}[tb]
  \centering
  \setlength{\tabcolsep}{12pt}
  \begin{tabular}{ll}
    (a) & (b) \vspace{-1.5em}\\
  \hspace{1.5em}\includegraphics[width=0.41\textwidth]{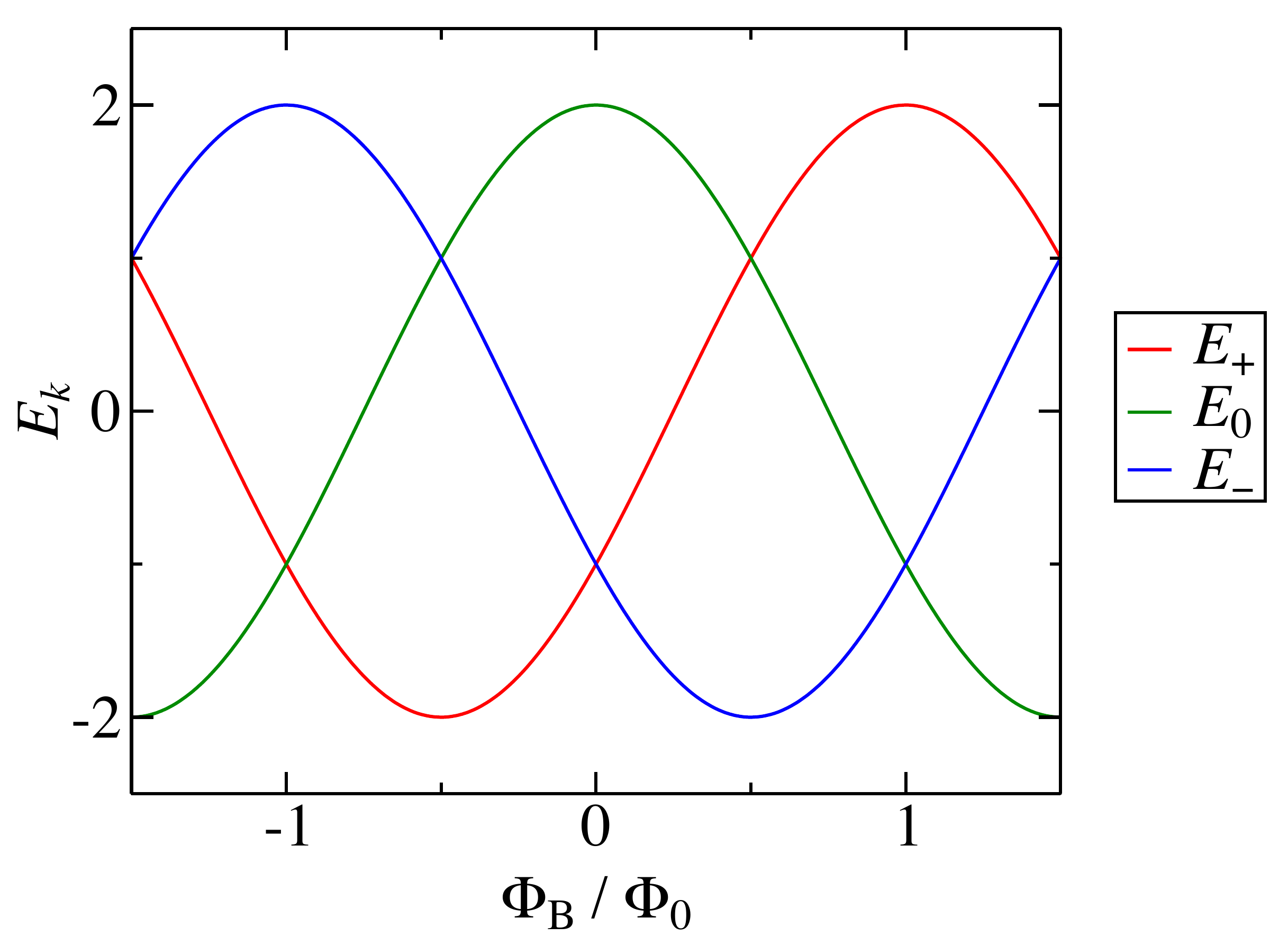} &
  \hspace{1.5em}\includegraphics[width=0.41\textwidth]{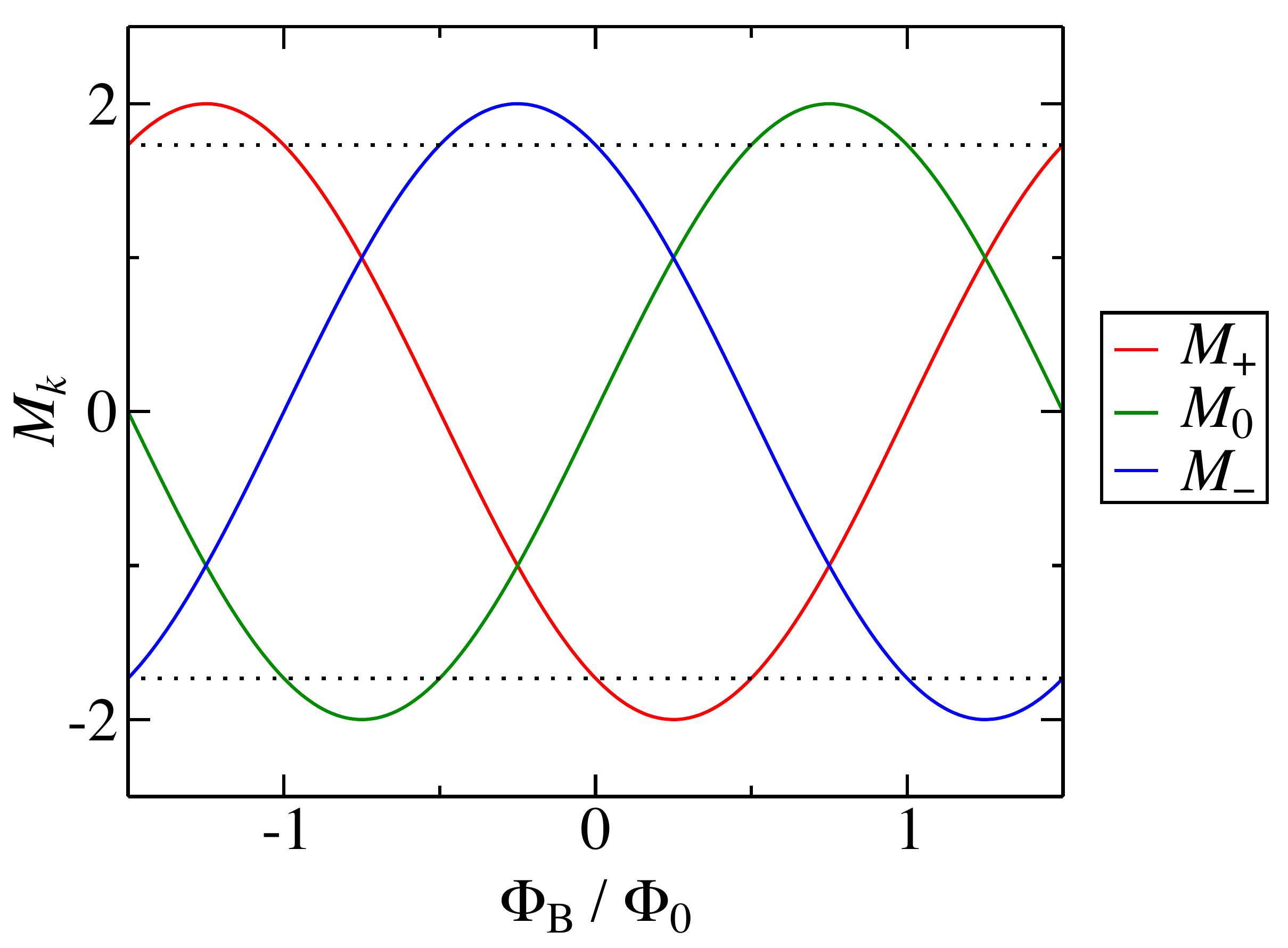}
  \end{tabular}
  \caption{\label{fig:model1} Properties of the trimer model with one orbital per site and no spin dependence, Eq.~\eqref{eq:model1}, as a function of the relative magnetic flux.
  Here $\NORM{t} = 1$.
  (a) Eigenenergies, Eq.~\eqref{eq:energies_model1}.
  (b) Orbital magnetic moment of each eigenstate, Eq.~\eqref{eq:magmoms_model1}.
  The dotted lines indicate the values $\pm\sqrt{3}$, the orbital moment for $k=\mp1$ when $B = 0$.}
\end{figure}

The eigenvalues and the corresponding orbital moments are plotted in Fig.~\ref{fig:model1}, as a function of the relative magnetic flux.
Both quantities are simple periodic functions of the magnetic flux.
Whenever two eigenstates are degenerate in energy, see Fig.~\ref{fig:model1}(a), their orbital moments are equal in magnitude and opposite in sign, cancelling each other, while the third eigenstate (which is non-degenerate) has zero orbital moment, as seen in Fig.~\ref{fig:model1}(b).
Thus, for an arbitrary electron filling in thermal equilibrium ($0 < N_{\MR{e}} < 3$), a finite net orbital moment requires lifting of the energy degeneracy.

For nanosized trimers and for realistic laboratory magnetic fields, $\Phi_{\MR{B}}/\Phi_0 \ll 1$, so the previous discussion might seem fanciful.
However, once spin exchange to a noncollinear magnetic structure is considered, such seemingly unrealistic effective magnetic fields do emerge.
We analyze this case in the next section.

\section{Noncollinear magnetism}\label{sec:model2}
We now extend the model from the previous section by including the spin exchange coupling,
\begin{equation}\label{eq:model2}
  \MC{H} = \MC{H}_{\MR{kin}} + \MC{H}_{\MR{mag}}
  = \sum_{i,j \neq i}\sum_s c_{is}^\dagger\,t_{ij}\,c_{js} + J\sum_i\sum_{ss'}c_{is}^\dagger\,\VEC{n}_i \cdot \paulivec_{ss'}\,c_{is'} \quad,
\end{equation}
with $s = \pm1$ the spin projection on an arbitrary quantization axis.
We refer to spin-up and spin-down by the associations $\uparrow\;= +1$ and $\downarrow\;= -1$.
In tandem with the orbital impact of the magnetic field, accounted for by the complex hopping parameters $t_{ij}$, the standard spin Zeeman coupling will also be considered.

The atomic structure is invariant under $2\pi/3$ rotations in real space around the central axis of symmetry of the triangle.
We similarly require the atomic plus magnetic structure to be invariant under the combination of a spatial rotation and a spin rotation, both by an angle of $2\pi/3$, around the respective rotation axes.
The local direction of the magnetic structure is thus chosen to be
\begin{equation}\label{eq:magdir}
  \VEC{n}_i =  \sin\theta\left(\cos\varphi_i\,\VEC{n}_x + \sin\varphi_i\,\VEC{n}_y\right) + \cos\theta\,\VEC{n}_z \quad,
\end{equation}
in spherical coordinates with respect to the spin quantization axis $\VEC{n}_z$, and the azimuthal angles are $\varphi_1 = 0$, $\varphi_2 = 2\pi/3$, and $\varphi_3 = -2\pi/3$, see Fig.~\ref{fig:trimer}(b).
We single out the following magnetic structures: ferromagnetic pointing along $+z$ (F$\uparrow$); planar triangular N\'eel structure (N); and ferromagnetic pointing along $-z$ (F$\downarrow$).
For these magnetic structures the scalar spin chirality takes the form
\begin{equation}\label{eq:schir}
   \VEC{n}_1 \cdot (\VEC{n}_2 \times \VEC{n}_3) = \frac{3\sqrt{3}}{2} \sin^2\theta \cos\theta = \frac{3\sqrt{3}}{2}\,C(\theta) \quad .
\end{equation}
This quantity is expected to play an important role as a driver of orbital magnetism.~\cite{Shindou2001,Tatara2002,Tatara2003,Tatara2003a,Nakamura2003a,Bulaevskii2008,Hoffmann2015,Dias2016}

We take as basis states the tensor product of the ring states from Eq.~\eqref{eq:eigenstates_model1} with the spin-up and spin-down eigenstates of $\sigma_z$:
\begin{equation}
  \ket{k\,s} = \frac{1}{\sqrt{3}} \left(\ket{1s} + e^{\iu\frac{2\pi k}{3}} \ket{2s} + e^{-\iu\frac{2\pi k}{3}} \ket{3s}\right) \quad.
\end{equation}
The basis states are not invariant under either reversal of translational motion, $\MC{T}_{\MR{P}} \ket{k\,s} = \ket{-k\,s}$, or of spin angular motion, $\MC{T}_{\MR{S}} \ket{k\,s} = -s\ket{k\,{-s}}$.
Defining $J_\perp = J\sin\theta$ and $J_z = J\cos\theta$, the spin exchange coupling is then expressed in this basis as
\begin{equation}\label{eq:spintriangle}
  \MC{H}_{\MR{mag}} = J\sum_i\VEC{n}_i \cdot \paulivec
  = J_\perp\big(\MC{R}_-\,\sigma_+ + \MC{R}_+\,\sigma_-\big) + J_z\,\sigma_z \quad,
\end{equation}
with the spin raising and lowering operators $\sigma_\pm = (\sigma_x \pm \iu\,\sigma_y)/2$, and the ring momentum raising and lowering operators defined in Eq.~\eqref{eq:kpm}.
Eq.~\eqref{eq:spintriangle} shows that the spin-flip part of the magnetic coupling exchanges spin angular momentum with ring momentum.
When the spin on every site is decreased ($\sigma_-$), the ring momentum $k$ increases by one unit ($\MC{R}_+$), and vice-versa.
This is the spin-orbit interaction driven by the noncollinear structure, in the present model.

The basis states couple pairwise, forming the hamiltonian blocks:
$\MC{H}_{\MR{a}}$ pairs $\ket{{-}1\,{\uparrow}}$ with $\ket{0\,{\downarrow}}$; $\MC{H}_{\MR{b}}$ pairs $\ket{0\,{\uparrow}}$ with $\ket{{+}1\,{\downarrow}}$; and $\MC{H}_{\MR{c}}$ pairs $\ket{{+}1\,{\uparrow}}$ pairs with $\ket{{-}1\,{\downarrow}}$.
Their matrix elements are
\begin{equation}
  \renewcommand{\arraystretch}{1.2}
  \begin{array}{r | c c}
    \MC{H}_\xi & \ket{k{-}1\,{\uparrow}} & \ket{k\,{\downarrow}} \\
    \hline
    \bra{k{-}1\,{\uparrow}} & E_{k{-}1} + J_z + B & J_\perp \\
    \bra{k\,{\downarrow}} & J_\perp & E_{k} - J_z - B
  \end{array} \quad,\quad \xi \in \{\MR{a},\MR{b},\MR{c}\} \quad ,
\end{equation}
where $E_k$ are the eigenenergies defined in Eq.~\eqref{eq:energies_model1}, and the spin Zeeman coupling to the external magnetic field was included, $B\,\sigma_z$.
Each block has then the eigenenergies (see Appendix~\ref{app:spindiag} and Eq.~\eqref{eq:ekprops})
\begin{equation}
  E_{\xi} = -\frac{E_{k{+}1}}{2} \pm \sqrt{\frac{3}{4}\,\big(M_{k{+}1}\big)^2 + \sqrt{3}\,\big(J_z + B\big)\,M_{k{+}1} + \big(J_z + B\big)^2 + J_\perp^2} \quad .
\end{equation}
For $\alpha = B = 0$ and introducing $x = \frac{3\NORM{t}}{2J}$, this yields
\begin{equation}\label{eq:energies_model2}
  E_{\MR{a}\pm} = \frac{\NORM{t}}{2} \pm J\sqrt{1 - 2\,x\cos\theta + x^2} \quad,\qquad
  E_{\MR{b}\pm} = \frac{\NORM{t}}{2} \pm J\sqrt{1 + 2\,x\cos\theta + x^2} \quad,\qquad
  E_{\MR{c}\pm} = -\NORM{t} \pm J \quad .
\end{equation}
The competition between kinetic and magnetic energies is encoded in the parameter $x$.

The magnetic moment associated with each of the eigenstates can be calculated directly from the eigenenergies.
It has two contributions (treating $B$ and $\alpha$ as independent):
\begin{equation}
  M_{\xi} = \frac{\partial E_{\xi}}{\partial B} -\frac{\partial E_{\xi}}{\partial\alpha}
  = M_{\MR{S}\xi} + M_{\MR{P}\xi} \quad .
\end{equation}
$M_{\MR{S}\xi}$ is the spin magnetic moment, arising from the Zeeman interaction, and signals the broken symmetry under reversal of spin angular momentum ($\MC{T}_{\MR{S}}$).
$M_{\MR{P}\xi}$ arises from the broken symmetry under reversal of the translational motion ($\MC{T}_{\MR{P}}$), due to the currents flowing around the trimer.
This is the orbital magnetic moment already encountered in the previous section.

For $\alpha = B = 0$ we find the spin moments:
\begin{subequations}\label{eq:magmoms_model2s}
\begin{align}
  M_{\MR{Sa}\pm} &= \pm \frac{\cos\theta - x}{\sqrt{1 - 2\,x\cos\theta + x^2}}
  \renewcommand{\arraystretch}{1.5}
  \approx \left\{ \begin{array}{l l}
  \pm\,\big(\cos\theta - x \sin^2\theta - \frac{3x^2}{2}\,C(\theta)\big) &,\qquad x \ll 1 \\
  \pm\,\big({-1} + \frac{1}{2x^2} \sin^2\theta + \frac{1}{x^3}\,C(\theta)\big) &,\qquad x \gg 1
  \end{array} \right. \quad , \\
  M_{\MR{Sb}\pm} &= \pm \frac{\cos\theta + x}{\sqrt{1 + 2\,x\cos\theta + x^2}}
  \renewcommand{\arraystretch}{1.5}
  \approx \left\{ \begin{array}{l l}
  \pm\,\big(\cos\theta + x \sin^2\theta - \frac{3x^2}{2}\,C(\theta)\big)  &,\qquad x \ll 1 \\
  \pm\,\big({+1} - \frac{1}{2x^2} \sin^2\theta + \frac{1}{x^3}\,C(\theta)\big) &,\qquad x \gg 1
  \end{array} \right. \quad , \\
  M_{\MR{Sc}\pm} &= \pm \cos\theta \quad .
\end{align}
\end{subequations}
$M_{\MR{S}}$ tells us about the spin character of an eigenstate.
A positive sign indicates $\uparrow$, a negative one $\downarrow$, and if it vanishes it has an equal amount of each character.
The adiabatic approximation, valid for $J \gg \NORM{t}$, makes the electron spin collinear with the direction of the magnetic structure.
This would lead to a $\cos\theta$ dependence, which is the first term in the $x \ll 1$ expansion.
The expansions were carried out up to the first term with the angular dependence of the scalar chirality, Eq.~\eqref{eq:schir}.

The orbital moments can then be shown to be simply related to the spin moments:
\begin{equation}\label{eq:magmoms_model2p}
  M_{\MR{Pa}} = M_{\MR{max}}\,\frac{M_{\MR{Sa}} + 1}{2} \quad , \qquad
  M_{\MR{Pb}} = M_{\MR{max}}\,\frac{M_{\MR{Sb}} - 1}{2} \quad , \qquad
  M_{\MR{Pc}} = -M_{\MR{max}}\,M_{\MR{Sc}} \quad ,
\end{equation}
with $M_{\MR{max}} = \sqrt{3}\,\NORM{t}$ the maximum value of the orbital magnetic moment for this model.

\begin{figure}[t]
  \centering
  \setlength{\tabcolsep}{2pt}
  \begin{tabular}{lll}
    (a) & (b) & (c) \vspace{-1.5em}\\
  \hspace{1.2em}\includegraphics[width=0.3\textwidth]{{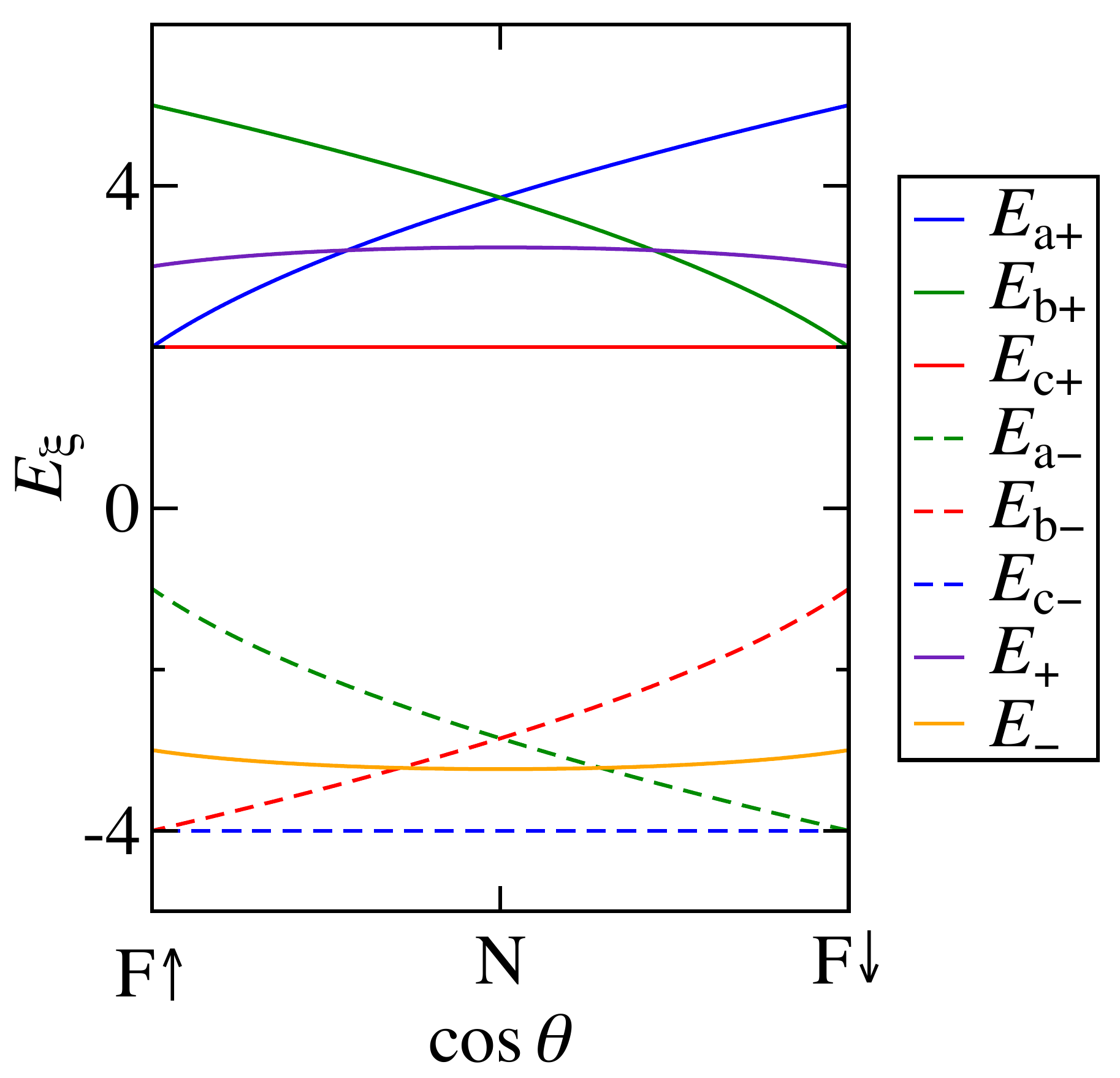}} &
  \hspace{1.3em}\includegraphics[width=0.3\textwidth]{{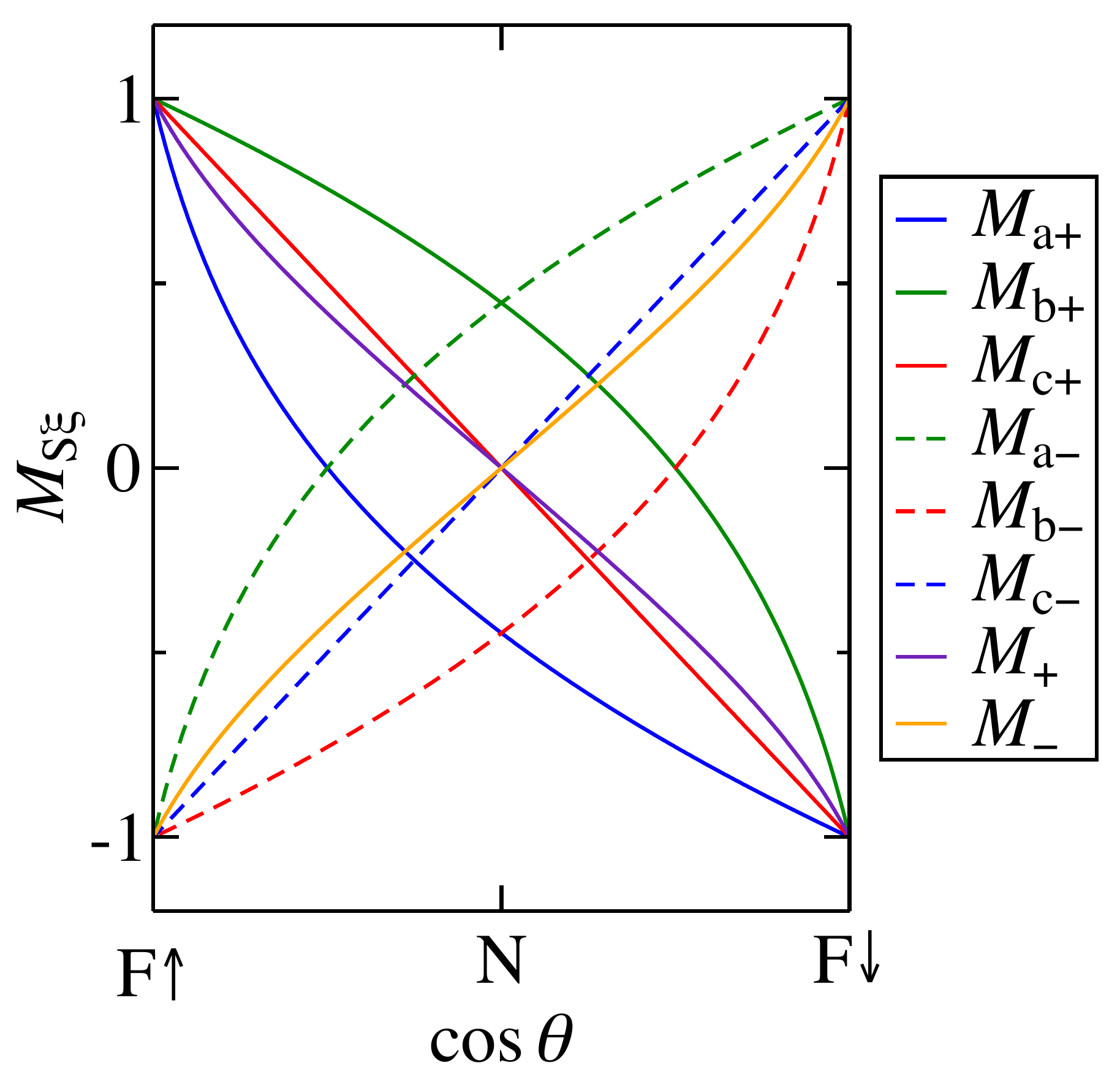}} &
  \hspace{1.2em}\includegraphics[width=0.3\textwidth]{{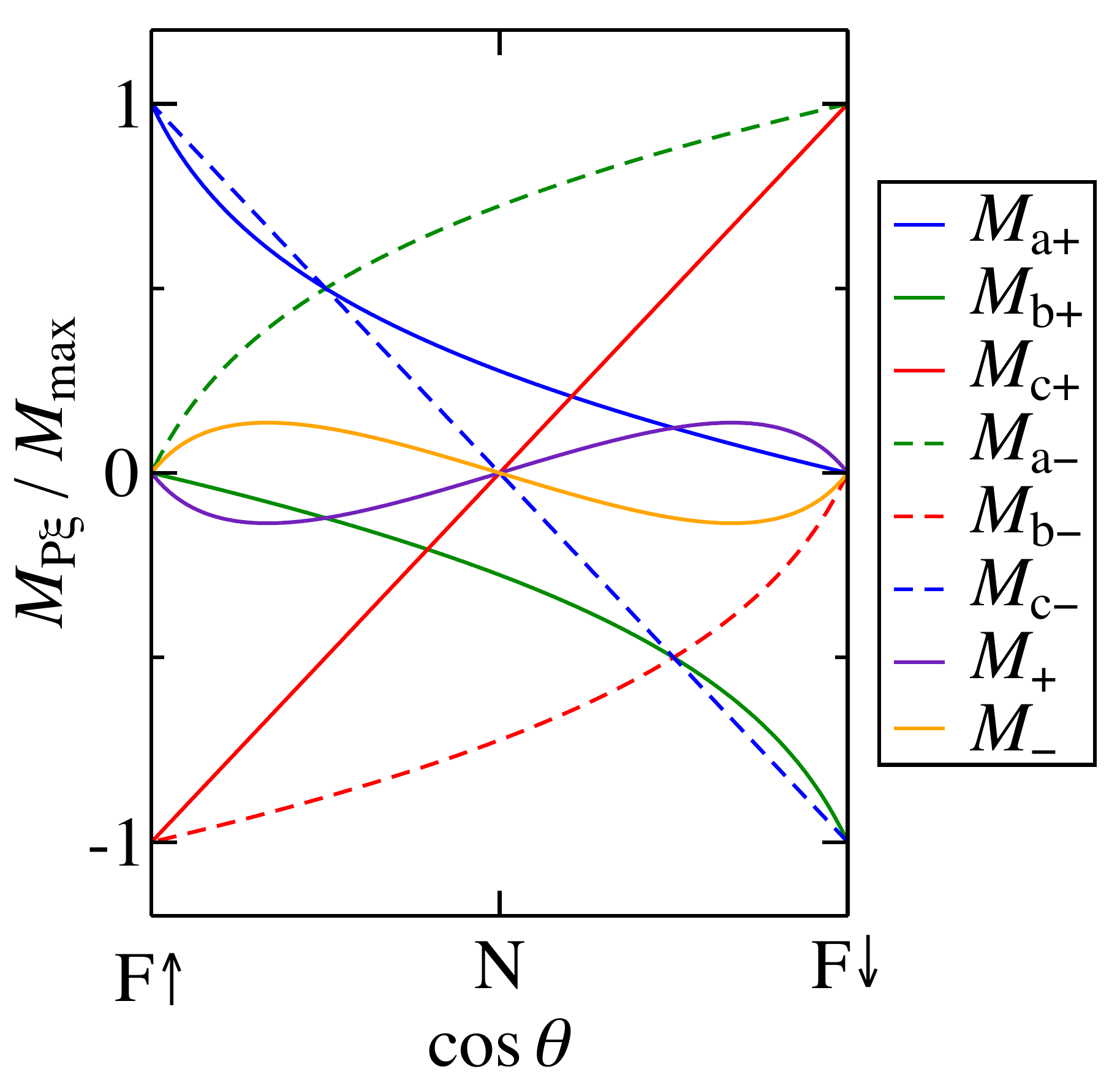}}
  \end{tabular}
  \caption{\label{fig:model2_J5} Trimer with one orbital per site and a noncollinear magnetic structure, Eq.~\eqref{eq:model2}, in the strong exchange coupling regime, $t = 1$ and $J = 3$ ($x = 1/2$).
  (a) Eigenenergies, Eq.~\eqref{eq:energies_model2}.
  (b) Spin magnetic moment of each eigenstate, Eq.~\eqref{eq:magmoms_model2s}.
  (c) Orbital magnetic moment of each eigenstate, Eq.~\eqref{eq:magmoms_model2p}.
  The magnetic structures are defined by Eq.~\eqref{eq:magdir} and illustrated in Fig.~\ref{fig:trimer}(b).
  The curves are labelled with the states for F$\uparrow$ taken as reference: the color labels the value of $M_{\MR{P}}$; solid lines and dashed lines indicate the sign of $M_{\MR{S}}$.
  The following combinations of eigenstates are also plotted: $(\lambda) = (\MR{a}\lambda) + (\MR{b}\lambda) + (\MR{c}\lambda)$, with $\lambda=\pm$.}
\end{figure}

First let us consider the case of the magnetic exchange dominating the kinetic energy, i.e.~$J \gg \NORM{t}$, allowing a comparison with the adiabatic approximation.
Fig.~\ref{fig:model2_J5} displays the results for $t = 1$ and $J = 3$ ($x = 1/2$).
The eigenenergies form two groups, separated by the exchange splitting $2J$, as seen in Fig.~\ref{fig:model2_J5}(a).
The magnetic noncollinearity effectively reduces the kinetic energy, evidenced by the shrinking `bandwidth' of each group when going from the F$\uparrow$ structure to the N structure.
Fig.~\ref{fig:model2_J5}(b) shows the spin magnetic moments.
The spin moments for the eigenstates labelled (c) follow perfectly the adiabatic approximation, seen as the linear behavior, while those for the eigenstates labelled (a) and (b) show deviations from the linear behavior.
Fig.~\ref{fig:model2_J5}(c) shows the orbital magnetic moments.
For the F$\uparrow$ structure, the eigenstates for each spin projection are decoupled and are just the ring states previously discussed, with the same orbital moments.
The variation of the orbital moments with the magnetic structure reveals the presence of the emergent magnetic field that it generates.
Going from F$\uparrow$ to N, we arrive at a new energy degeneracy.
Comparison of the evolution of the curves with those in Fig.~\ref{fig:model1} lets us assign $\Phi_{\MR{B}}/\Phi_0 = -1/2$ to the eigenstates evolving from spin-up, and $\Phi_{\MR{B}}/\Phi_0 = +1/2$ for those evolving from spin-down.
The net orbital moment is zero for the ferromagnetic structures, but not for the noncollinear ones.
To illustrate this, we sum all the contributions corresponding to the $+$ and $-$ bands, which corresponds to placing three electrons in the three upper or lower eigenstates, see Fig.~\ref{fig:model2_J5}(a).
The average spin for these combinations follows the magnetic structure almost linearly, see Fig.~\ref{fig:model2_J5}(b), the behavior expected in the adiabatic limit.
From Fig.~\ref{fig:model2_J5}(c) we observe that the average orbital moments are indeed zero for the F endpoints and for the N structure, but are finite for the noncollinear structures.
A comparison with the $x \ll 1$ expansion in Eq.~\ref{eq:magmoms_model2s} shows that $M_{\MR{P}\pm} \propto C(\theta)$, the scalar spin chirality, to leading order.

\begin{figure}[t]
  \centering
  \setlength{\tabcolsep}{2pt}
  \begin{tabular}{lll}
    (a) & (b) & (c) \vspace{-1.5em}\\
  \hspace{1.2em}\includegraphics[width=0.3\textwidth]{{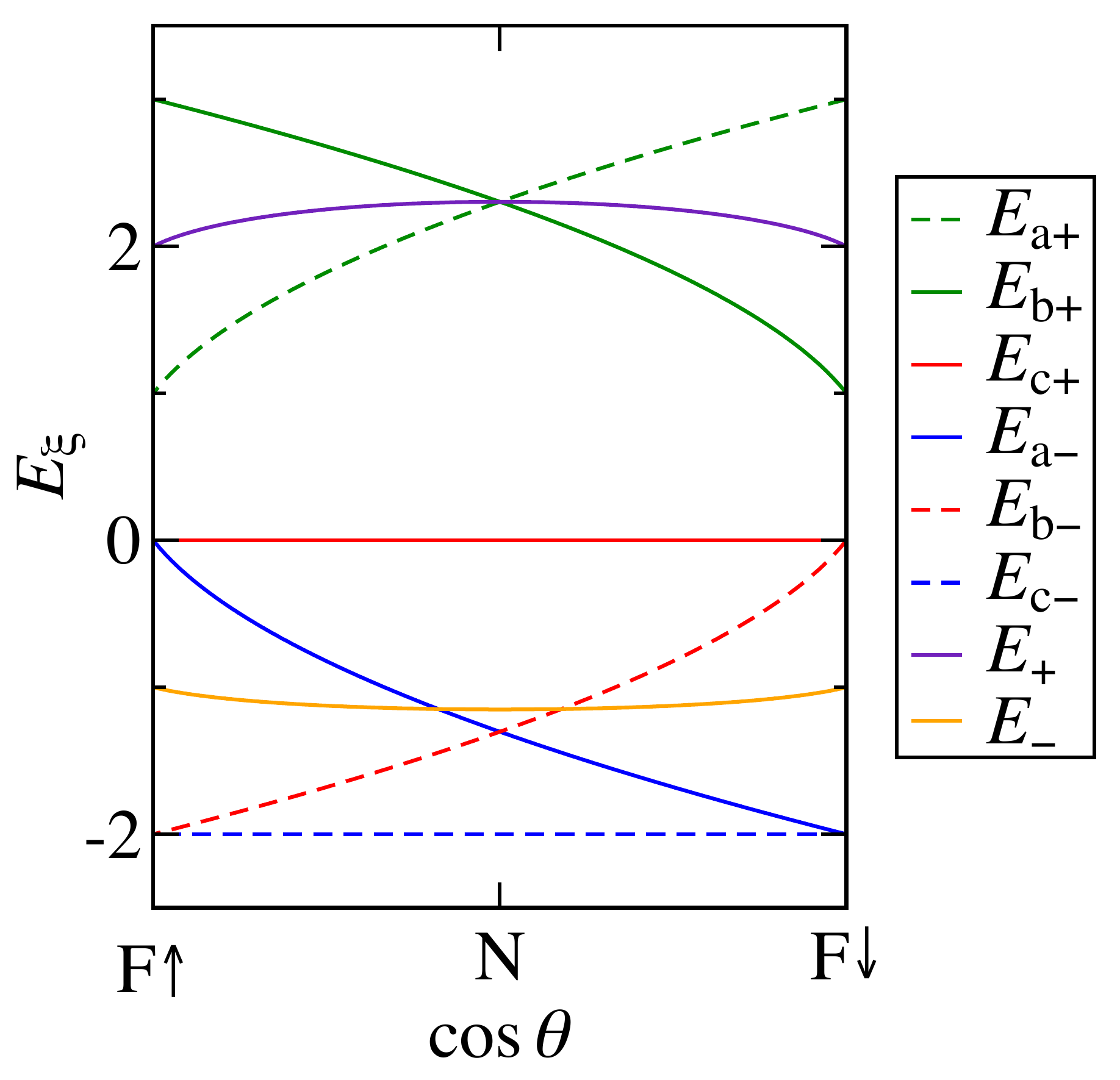}} &
  \hspace{1.3em}\includegraphics[width=0.3\textwidth]{{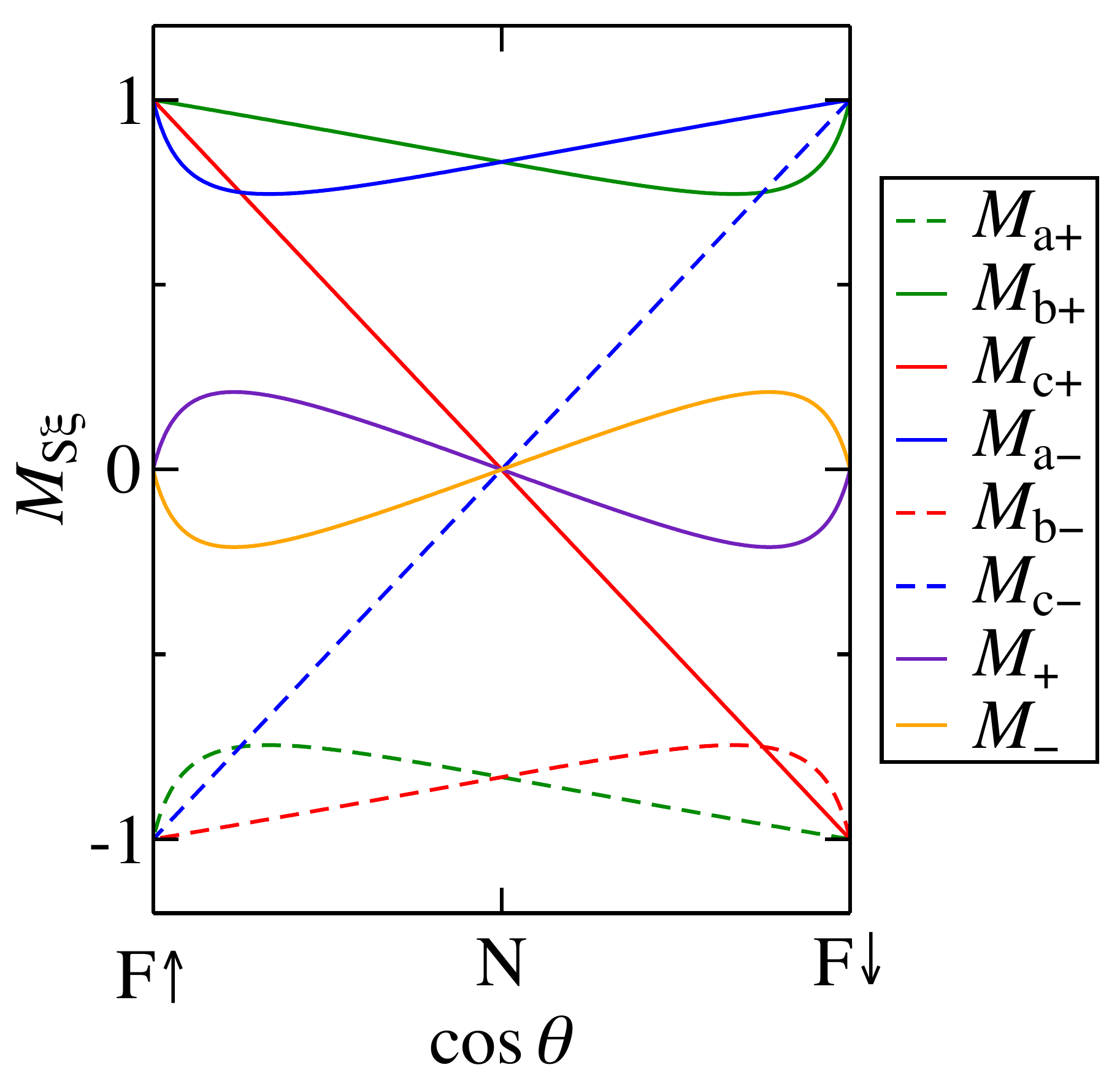}} &
  \hspace{1.2em}\includegraphics[width=0.3\textwidth]{{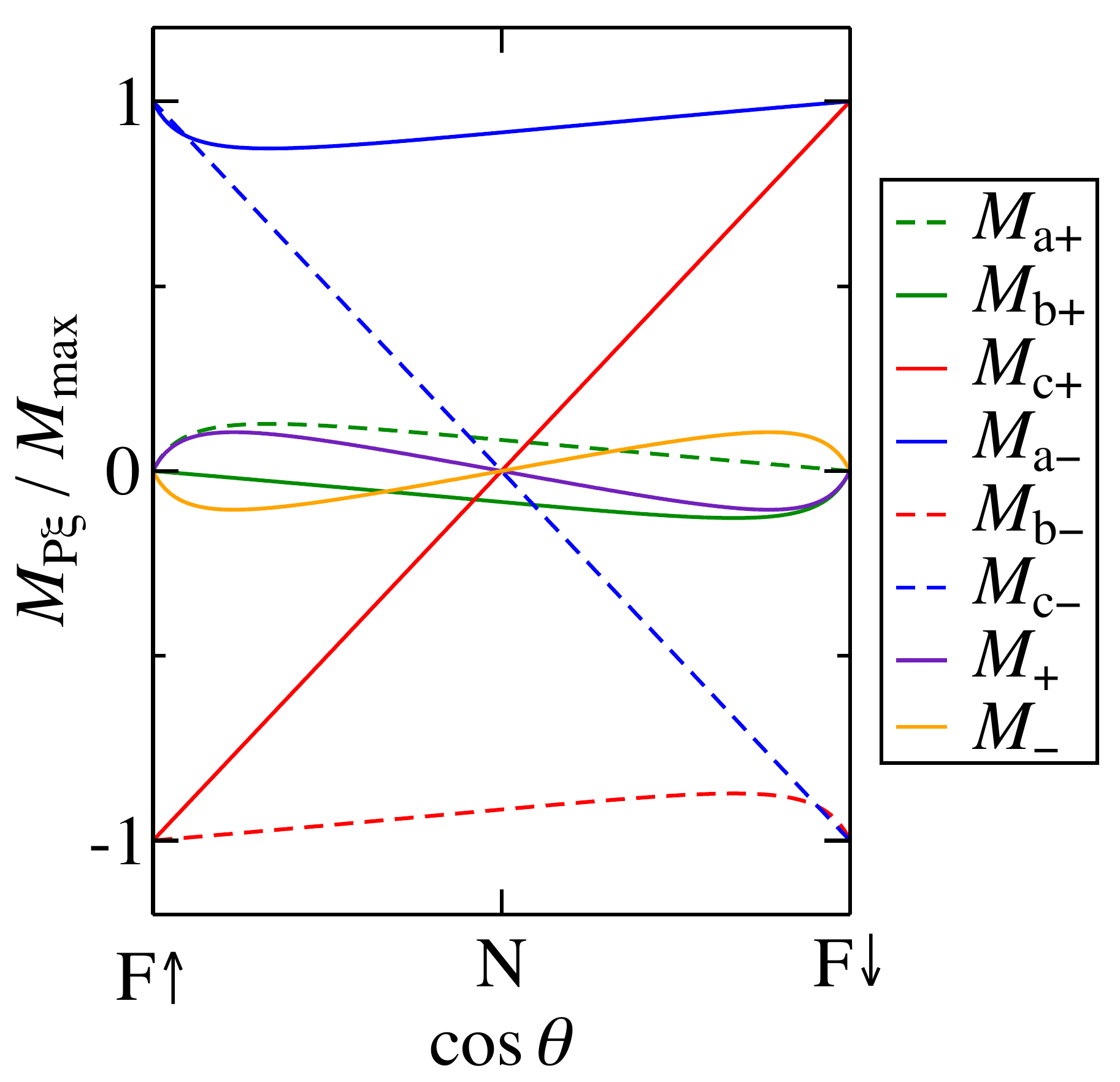}}
  \end{tabular}
  \caption{\label{fig:model2_J1} Trimer with one orbital per site and a noncollinear magnetic structure, Eq.~\eqref{eq:model2}, in the weak exchange coupling regime, $t = 1$ and $J = 1$ ($x = 3/2$).
  (a) Eigenenergies, Eq.~\eqref{eq:energies_model2}.
  (b) Spin magnetic moment of each eigenstate, Eq.~\eqref{eq:magmoms_model2s}.
  (c) Orbital magnetic moment of each eigenstate, Eq.~\eqref{eq:magmoms_model2p}.
  The magnetic structures are defined by Eq.~\eqref{eq:magdir} and illustrated in Fig.~\ref{fig:trimer}(b).
  The curves are labelled with the states for F$\uparrow$ taken as reference: the color labels the value of $M_{\MR{P}}$; solid lines and dashed lines indicate the sign of $M_{\MR{S}}$.
  The following combinations of eigenstates are also plotted: $(+) = (\MR{c}+) + (\MR{c}-)$ and $(-) = (\MR{a}+) + (\MR{a}-) + (\MR{b}+) + (\MR{b}-)$.}
\end{figure}

Next consider the case of the magnetic exchange being comparable to the kinetic energy, i.e.~$J \sim \NORM{t}$.
Fig.~\ref{fig:model2_J1} displays the results for $t = 1$ and $J = 1$ ($x = 3/2$).
As $2J < 3t$, the eigenergies for spin-up and spin-down overlap, see Fig.~\ref{fig:model2_J1}(a).
Comparing with the previous case, we see that the ordering of the states for F$\uparrow$ has changed, as indicated by the sequence of colors in the figure.
This has a dramatic impact on the behavior of the magnetic moments, Fig.~\ref{fig:model2_J1}(b,c).
On the one hand, the eigenstates labelled (c) still follow the linear behavior.
On the other hand, both the spin and the orbital moments for the eigenstates labelled (a) and (b) are only weakly modified by the magnetic structure.
This has a simple explanation: the ring states coupled in $\MC{H}_{\MR{c}}$ are degenerate in energy for $J = 0$, and so the coupling to the magnetic structure is always non-perturbative, while $\MC{H}_{\MR{a}}$ and $\MC{H}_{\MR{b}}$ each couple states split by $3t$ for $J = 0$, and so the exchange coupling has only a perturbative effect.
To visualize whether there is a net orbital moment also in this case, we sum all the contributions corresponding either to the four lower or to the two higher eigenstates, see Fig.~\ref{fig:model2_J1}(a).
Although the net spin moment is zero for the F and N structures, surprisingly it acquires a finite value for the noncollinear structures, see Fig.~\ref{fig:model2_J1}(b).
Fig.~\ref{fig:model2_J1}(c) shows that the net orbital moment has a similar behavior.
A comparison with the $x \gg 1$ expansion in Eq.~\ref{eq:magmoms_model2s} shows that $M_{\MR{P}\pm} \propto M_{\MR{S}\pm} \propto C(\theta)$, the scalar spin chirality, to leading order.

We have thus seen how a noncollinear magnetic structure can lead to orbital magnetic effects in the absence of the RSOI.
The impact on the electronic structure depends crucially on whether the states which become coupled by the magnetic exchange are initially degenerate in energy or not.
For the former the adiabatic approximation is always valid, while for the latter the exchange coupling must overcome the difference in kinetic energy between the states for it to have a strong influence.
The picture is also very difference if each eigenstate is considered by itself or if a group of eigenstates is considered together.
In the next section the RSOI is introduced in the model, and its consequences analyzed.

\section{Interplay between noncollinear magnetism and the relativistic spin-orbit interaction}\label{sec:model3}
We extend our model one final time, by taking the orbital character of the electrons on every site into account.
Following Ref.~\onlinecite{Dias2016}, we consider two $d$-orbitals to be present on every site, namely $\ket{xy}$ and $\ket{x^2{-}y^2}$, assumed to be initially degenerate in energy.
We shall work with their complex counterparts, which are eigenstates of $L_z$:
\begin{equation}
  \ket{\pm} = \frac{1}{\sqrt{2}}\left(\ket{x^2{-}y^2} \pm \iu \ket{xy}\right) \quad,\qquad L_z \ket{\pm2} = \pm2 \ket{\pm2} \quad .
\end{equation}
The RSOI in atomic form reduces to $\VEC{L}\cdot\bm{\upsigma} = L_z\,\sigma_z$, as the other angular momentum operators vanish when restriced to these two orbitals.
To label the states, we make the identifications $+2 = {\circlearrowleft}$ and  $-2 = {\circlearrowright}$.

However, the kinetic hamiltonian now has to describe the directionality of the orbitals.
For zero magnetic field, the hopping matrix in either the real or complex basis is given by
\begin{equation}
  \renewcommand{\arraystretch}{1.2}
  \begin{array}{r|cc}
    t_{ij} & \ket{x^2{-}y^2} & \ket{xy} \\
    \hline
    \bra{x^2{-}y^2} & t \left(1 + \cos \gamma_{ij}\right) & t \sin \gamma_{ij} \\
    \bra{xy} & t \sin \gamma_{ij} & t \left(1 - \cos \gamma_{ij}\right)
  \end{array} \qquad\text{or}\qquad
  \begin{array}{r|cc}
    t_{ij} & \ket{\circlearrowleft} & \ket{\circlearrowright} \\
    \hline
    \bra{\circlearrowleft} & t  & t\,e^{-\iu\gamma_{ij}} \\
    \bra{\circlearrowright} & t\,e^{\iu\gamma_{ij}} & t
  \end{array} \quad,
\end{equation}
where $\gamma_{ij}/4$ is the angle between the bond and the $x$-axis.
We have $\gamma_{ij} = \gamma_{ji}$ (mod $2\pi$), and $\gamma_{12} = 0$, $\gamma_{23} = 2\pi/3$ and $\gamma_{31} = -2\pi/3$.
This encompasses the fourfold symmetry of the orbitals, and their directionality.
For example, if two orbitals are along the $x$-axis, hopping can only occur if they are both of $\ket{x^2{-}y^2}$ type.
We can encode the action of the hopping matrix on the orbitals using a new set of Pauli matrices $\tau_\mu$ (to be distinguished from the ones used for spin),
\begin{equation}
  t_{ij} = \NORM{t}\,e^{\iu\chi_{ij}\alpha} \big(\tau_0 + e^{-\iu\gamma_{ij}}\tau_+ + e^{\iu\gamma_{ij}}\tau_-\big) = t_{ji}^\dagger \quad,
\end{equation}
where the magnetic field was restored via the Peierls phase, see Eq.~\eqref{eq:tbpeierls}.

Our basis states are the tensor product of the ring states $k = 0,\pm1$, of the two orbitals $m = \pm2$, and of the spinors $s = \pm1$:
\begin{equation}
  \ket{k\,m\,s} = \frac{1}{\sqrt{3}} \left(\ket{1\,m\,s} + e^{\iu\frac{2\pi k}{3}} \ket{2\,m\,s} + e^{-\iu\frac{2\pi k}{3}} \ket{3\,m\,s}\right) \quad.
\end{equation}
These basis functions are ideal to describe time-reversal symmetry breaking: the time-reversed counterpart of each basis state is $\MC{T}\ket{k\,m\,s} = -s\ket{-k\,{-m}\,{-s}}$, which corresponds to reversing translational, orbital and spin motions, i.e., reversing each of the variables describing the state of motion.
The action of the hamiltonian can then be separated into a diagonal part (that leaves the basis state unchanged), and different kinds of off-diagonal terms, according to what change they effect on the basis state.

The diagonal part of the hamiltonian is
\begin{equation}
  \MC{H}_0 = t\,\MC{R}_0 + \bar{t}\,\MC{R}_0^\dagger + J_z\,\sigma_z + \xi\,\tau_z\,\sigma_z + B\,\tau_z \quad.
\end{equation}
The new terms are the RSOI ($\xi$ term), and the orbital Zeeman coupling ($B$ term).
This part of the hamiltonian acts on a basis state as, recalling~Eq.~\eqref{eq:energies_model1},
\begin{equation}
  \MC{H}_0 \ket{k\,m\,s} = \big(E_k + s\,J_z + \MR{sgn}(m)\,(s\,\xi + B)\big)\ket{k\,m\,s} \quad .
\end{equation}
We have already seen from the previous section that there are two terms that exchange spin angular momentum and ring momentum,
\begin{equation}
  \MC{H}_{\MR{S}\pm} = J_\perp \MC{R}_\mp\,\sigma_\pm \quad,\qquad
  \MC{H}_{\MR{S}\pm} \ket{k\,m\,s} = J_\perp \ket{k{\mp}1\,m\,s{\pm}1} \quad .
\end{equation}
These terms led to the spin-orbit interaction generated by the noncollinear magnetic structure.
The remaining piece of the kinetic term generates two terms that exchange orbital angular momentum and ring momentum,
\begin{equation}
  \MC{H}_{\MR{L}\pm} = \big(t\,\MC{R}_0\,\MC{R}_\mp + \bar{t}\,\MC{R}_\mp\,\MC{R}_0^\dagger\big)\,\tau_\pm \quad,
\end{equation}
recall Eq.~\eqref{eq:kpm}, with the result
\begin{align}
  \MC{H}_{\MR{L}\pm} \ket{k\,m\,s} = \Big(t\,e^{-\iu\frac{2\pi(k\mp1)}{3}} + \bar{t}\,e^{\iu\frac{2\pi k}{3}}\Big) \ket{k{\mp}1\,m{\pm}2\,s}
  = E_{k{\pm}1}\,e^{\mp\iu\frac{2\pi}{3}} \ket{k{\mp}1\,m{\pm}2\,s} \quad ,
\end{align}
according to the definition in Eq.~\eqref{eq:energies_model1}.
This might be called an orbit-orbit interaction, as the translational motion and the local orbital motion are coupled.

Our hamiltonian can now be written as initially presented in Eq.~\eqref{eq:tbmodel},
\begin{equation}
  \MC{H} = \MC{H}_{\MR{kin}} + \MC{H}_{\MR{mag}} + \MC{H}_{\MR{soi}}
  = \MC{H}_0 + \MC{H}_{\MR{S}+} + \MC{H}_{\MR{S}-} + \MC{H}_{\MR{L}+} + \MC{H}_{\MR{L}-} \quad ,
\end{equation}
and represents a $12 \times 12$ matrix, composed of three $4 \times 4$ blocks.
The basis states that can be coupled by the hamiltonian are limited by $(\MC{H}_{\MR{S}\pm})^2 \ket{k\,m\,s} = 0$ and $(\MC{H}_{\MR{L}\pm})^2 \ket{k\,m\,s} = 0$, as the spin and the atomic angular momentum cannot be raised or lowered more than once.
We then have the following chain of coupled states:
\begin{equation}
  \ket{k{-}1\,{\circlearrowleft}\,{\uparrow}}
  \;\underset{\MC{H}_{\MR{L}-}}{\longrightarrow}\; \ket{k\,{\circlearrowright}\,{\uparrow}}
  \;\underset{\MC{H}_{\MR{S}-}}{\longrightarrow}\; \ket{k{+}1\,{\circlearrowright}\,{\downarrow}}
  \;\underset{\MC{H}_{\MR{L}+}}{\longrightarrow}\; \ket{k\,{\circlearrowleft}\,{\downarrow}}
  \;\underset{\MC{H}_{\MR{S}+}}{\longrightarrow}\; \ket{k{-}1\,{\circlearrowleft}\,{\uparrow}} \quad .
\end{equation}
The three blocks are generated by the three possible starting values of $k$, and can be organized as follows:
$\MC{H}_{\MR{a}}$ couples $\ket{0\,{\circlearrowleft}\,{\uparrow}}$, $\ket{{+}1\,{\circlearrowright}\,{\uparrow}}$, $\ket{{+}1\,{\circlearrowleft}\,{\downarrow}}$, and $\ket{{-}1\,{\circlearrowright}\,{\downarrow}}$;
$\MC{H}_{\MR{b}}$ couples $\ket{{+}1\,{\circlearrowleft}\,{\uparrow}}$, $\ket{{-}1\,{\circlearrowright}\,{\uparrow}}$, $\ket{{-}1\,{\circlearrowleft}\,{\downarrow}}$, and $\ket{0\,{\circlearrowright}\,{\downarrow}}$;
and $\MC{H}_{\MR{c}}$ couples $\ket{{-}1\,{\circlearrowleft}\,{\uparrow}}$, $\ket{0\,{\circlearrowright}\,{\uparrow}}$, $\ket{0\,{\circlearrowleft}\,{\downarrow}}$, and $\ket{{+}1\,{\circlearrowright}\,{\downarrow}}$.
The matrix elements for these blocks have the form
\begin{equation}\label{eq:hblock}
  \renewcommand{\arraystretch}{1.5}
  \begin{array}{r | c c | c c}
  \MC{H}_\xi & \ket{k{-}1\,{\circlearrowleft}\,{\uparrow}} & \ket{k\,{\circlearrowright}\,{\uparrow}} & \ket{k\,{\circlearrowleft}\,{\downarrow}} & \ket{k{+}1\,{\circlearrowright}\,{\downarrow}} \\
  \hline
  \bra{k{-}1\,{\circlearrowleft}\,{\uparrow}} & E_{k-1} + J_z + \xi + B & E_{k{+}1}\,e^{-\iu\frac{2\pi}{3}} & J_\perp & 0 \\
  \bra{k\,{\circlearrowright}\,{\uparrow}} & E_{k{+}1}\,e^{\iu\frac{2\pi}{3}} & E_k + J_z - \xi - B & 0 & J_\perp \\
  \hline
  \bra{k\,{\circlearrowleft}\,{\downarrow}} & J_\perp & 0 & E_k - J_z - \xi + B & E_{k{-}1}\,e^{-\iu\frac{2\pi}{3}} \\
  \bra{k{+}1\,{\circlearrowright}\,{\downarrow}} & 0 & J_\perp & E_{k{-}1}\,e^{\iu\frac{2\pi}{3}} & E_{k+1} - J_z + \xi - B
  \end{array} \quad .
\end{equation}

The case of a noncollinear magnetic structure is analytically cumbersome, as the hamiltonian blocks are $4 \times 4$ matrices.
If the magnetic exchange is much stronger than all the other terms, we can adopt the frequently used adiabatic approximation.~\cite{Sitte2014}
The spin projectors that diagonalize the magnetic exchange interaction are (see Appendix~\ref{app:spindiag})
\begin{equation}
  P_\pm = \frac{1}{2}\,\big(\sigma_0 \pm (\cos\theta\,\sigma_z + \sin\theta\,\sigma_x)\big) \quad .
\end{equation}
As the results for $s = -1$ can be obtained from those for $s = +1$ by the replacements $J \rightarrow -J$ and $\cos\theta \rightarrow -\cos\theta$, we set $s = +1$ and drop the spin label in the following.

Tracing over the spin components, we define an effective hamiltonian by
\begin{align}
  \widetilde{\MC{H}}_{\xi} = \braket{+|\MC{H}_\xi|+} = \Tr\,P_+\,\MC{H}_\xi = \frac{\MC{H}_{\xi,\uparrow\uparrow} + \MC{H}_{\xi,\downarrow\downarrow}}{2}
  + \Big(\!\cos\theta\,\frac{\MC{H}_{\xi,\uparrow\uparrow} - \MC{H}_{\xi,\downarrow\downarrow}}{2} + \sin\theta\,J_\perp\Big)\,\tau_0 \quad ,
\end{align}
which can be written using the orbital Pauli matrices as
\begin{align}\label{eq:hblocksp}
  \widetilde{\MC{H}}_{\xi} = J\,\tau_0 &+ \frac{1}{4}\,\Big(\big(E_{k} - \sqrt{3}\,M_k\cos\theta\big)\,\tau_0
  - \big(\sqrt{3}\,M_{k} + 3\,E_{k}\cos\theta - 4\,(\xi\cos\theta + B)\big)\,\tau_z\Big) \nonumber\\
  &- \frac{1}{2}\,\big(E_{k} - \sqrt{3}\,M_k\cos\theta\big) \big(e^{-\iu\frac{2\pi}{3}}\tau_+ + e^{\iu\frac{2\pi}{3}}\tau_-\big) \quad .
\end{align}
The only role played by $J$ is to define the energy zero, so we will also set $J = 0$ from now on.

The eigenergies for the effective hamiltonian blocks $\widetilde{\MC{H}}_{\xi}$ are 
\begin{align}
  E_{\xi} = \frac{1}{4}\,\big(E_{k} - \sqrt{3}\,M_k\cos\theta \pm \NORM{t} \sqrt{D_{\xi}}\,\big) \quad ,
\end{align}
with the discriminants
\begin{equation}
  \NORM{t}^2 D_{\xi} = \big(\sqrt{3}\,M_{k} + 3\,E_{k}\cos\theta - 4\,(\xi\cos\theta + B)\big)^2 + 4\,\big(E_{k} - \sqrt{3}\,M_k\cos\theta\big)^2 \quad .
\end{equation}
The orbital moments can be decomposed into two contributions (taking $B$ and $\alpha$ to be independent),
\begin{equation}
  M_{\xi} = \frac{\partial E_{\xi}}{\partial B} -\frac{\partial E_{\xi}}{\partial\alpha}
  = M_{\MR{L}\xi} + M_{\MR{P}\xi} \quad .
\end{equation}
$M_{\MR{L}\xi}$ is the atomic orbital moment, stemming from the orbital Zeeman interaction, and signals the broken symmetry under reversal of the local orbital motion ($\MC{T}_{\MR{L}}$).
In the previous section we already encountered $M_{\MR{P}\xi}$, the contribution to the orbital motion from the currents circulating around the trimer.
In the adiabatic approximation $M_{\MR{S}\xi} = \cos\theta$ by construction, so it does not merit further consideration.

For $\alpha = B = 0$ and setting $y = 4\,\xi/\NORM{t}$, we obtain the eigenergies
\begin{subequations}\label{eq:energies_model3}
\begin{equation}
  E_{\MR{a}\pm} = \frac{\NORM{t}}{4}\,\Big({-}1 + 3\cos\theta \pm \sqrt{13 - 6\,\big(1 - y\big)\cos\theta + \big(45 + 6\,y + y^2\big)\cos^2\theta}\,\Big) \quad ,
\end{equation}
\begin{equation}
  E_{\MR{b}\pm} = \frac{\NORM{t}}{4}\,\Big({-}1 - 3\cos\theta \pm \sqrt{13 + 6\,\big(1 - y\big)\cos\theta + \big(45 + 6\,y + y^2\big)\cos^2\theta}\,\Big) \quad ,
\end{equation}
\begin{equation}
  E_{\MR{c}\pm} = \frac{\NORM{t}}{4}\,\Big(2 \pm \sqrt{16 + \big(6 - y\big)^2\cos^2\theta}\,\Big) \quad .
\end{equation}
\end{subequations}

The atomic orbital moments are
\begin{equation}\label{eq:magmoms_model3l}
  M_{\MR{L}\MR{a}\pm} = \pm\frac{(y + 3) \cos\theta + 3}{\sqrt{D_{\MR{a}}}} \quad , \qquad
  M_{\MR{L}\MR{b}\pm} = \pm\frac{(y + 3) \cos\theta - 3}{\sqrt{D_{\MR{b}}}} \quad , \qquad
  M_{\MR{L}\MR{c}\pm} = \pm\frac{(y - 6) \cos\theta}{\sqrt{D_{\MR{c}}}} \quad .
\end{equation}
$M_{\MR{L}}$ tells us about the atomic orbital character of an eigenstate.
A positive sign indicates $\circlearrowleft$, a negative one $\circlearrowright$, and if it vanishes it has an equal amount of each character.

The orbital moments arising from the circulating currents are (recall $M_{\MR{max}} = \sqrt{3}\,\NORM{t}$)
\begin{subequations}\label{eq:magmoms_model3p}
\begin{equation}
  M_{\MR{P}\MR{a}\pm} = -\frac{M_{\MR{max}}}{4} \left(\cos\theta + 1 \pm \frac{-1 + (2 + y)\cos\theta + 3\,(1 - y)\cos^2\theta}{\sqrt{D_{\MR{a}}}}\right) \quad ,
  \end{equation}
\begin{equation}
  M_{\MR{P}\MR{b}\pm} = -\frac{M_{\MR{max}}}{4} \left(\cos\theta - 1 \pm \frac{+1 + (2 + y)\cos\theta - 3\,(1 - y)\cos^2\theta}{\sqrt{D_{\MR{b}}}}\right) \quad ,
\end{equation}
\begin{equation}
  M_{\MR{P}\MR{c}\pm} = \frac{M_{\MR{max}}}{2} \left(1 \pm \frac{2 + y}{\sqrt{D_{\MR{c}}}}\right) \cos\theta \quad .
\end{equation}
\end{subequations}
They can be used to characterize the translational motion, as in Sec.~\ref{sec:model1}.
For a ferromagnetic structure, $M_{\MR{P}}/M_{\MR{max}} \approx \pm 1$ can be associated with $k = \mp1$, and $M_{\MR{P}}/M_{\MR{max}} \approx 0$ with $k=0$.
Note that there is no simple relation between $M_{\MR{L}}$ and $M_{\MR{P}}$, in contrast to the results of the previous section.

\begin{figure}[t]
  \centering
  \setlength{\tabcolsep}{2pt}
  \begin{tabular}{lll}
    (a) & (b) & (c) \vspace{-1.5em}\\
  \hspace{1.2em}\includegraphics[width=0.3\textwidth]{{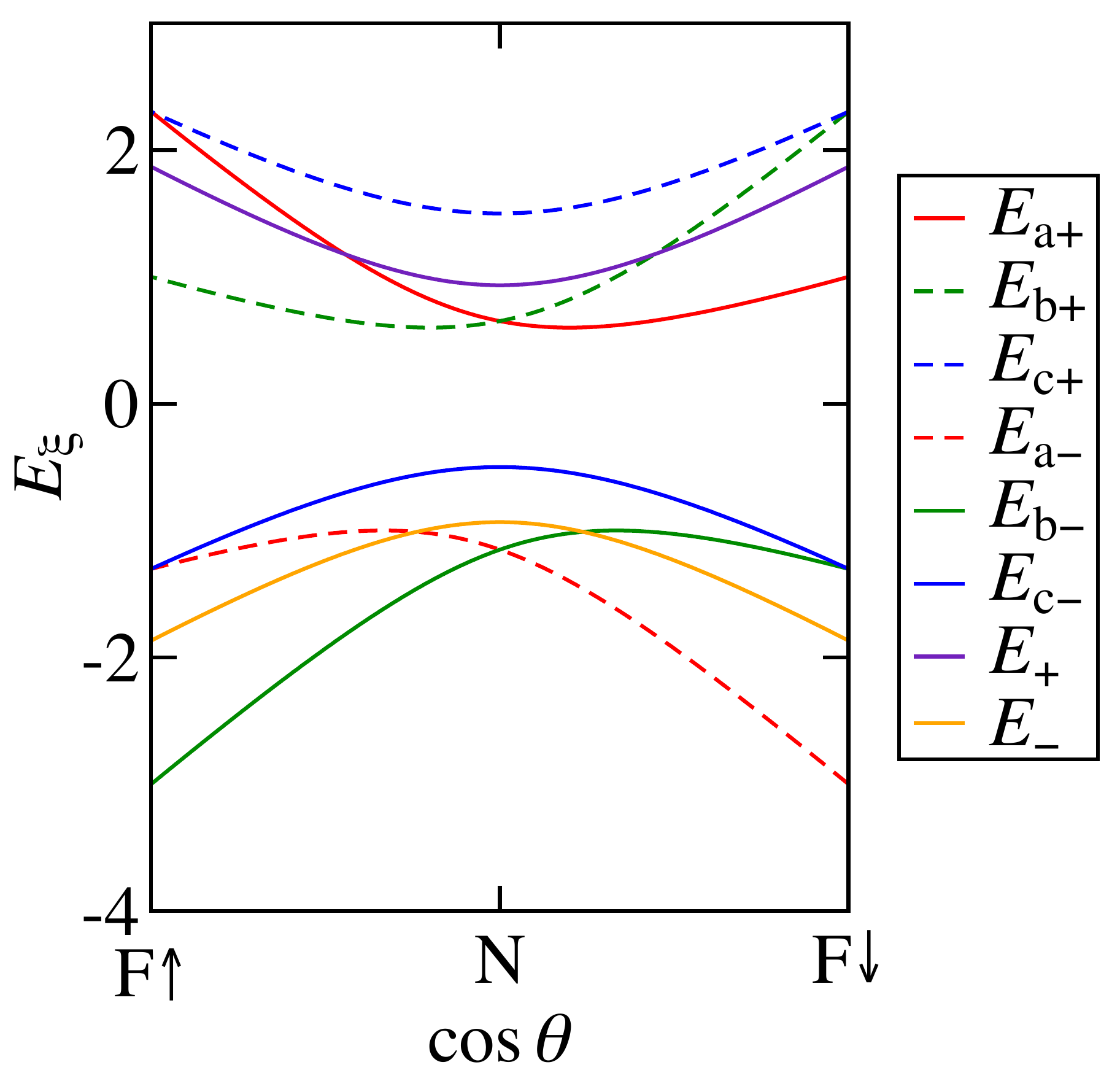}} &
  \hspace{1.3em}\includegraphics[width=0.3\textwidth]{{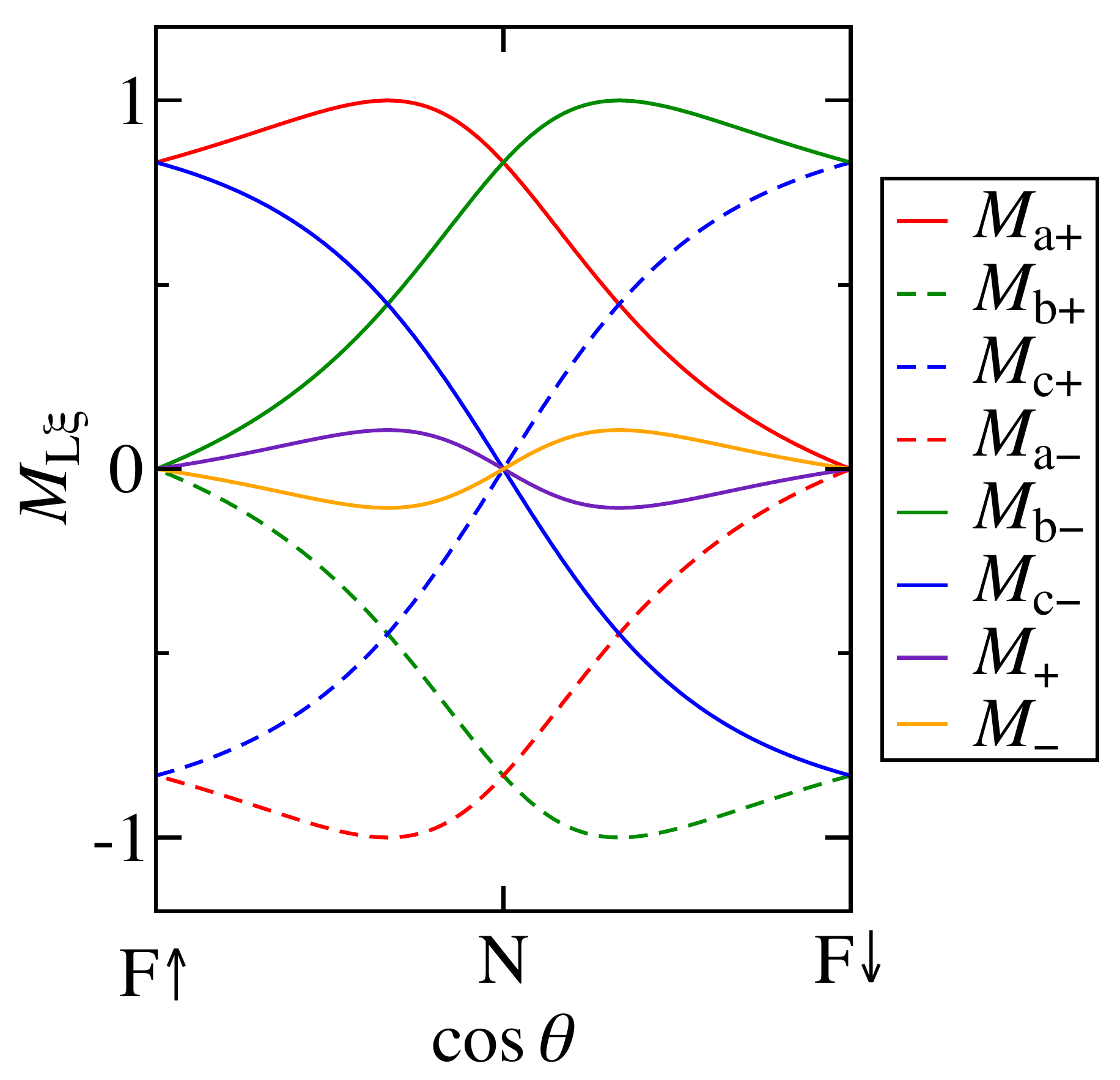}} &
  \hspace{1.2em}\includegraphics[width=0.3\textwidth]{{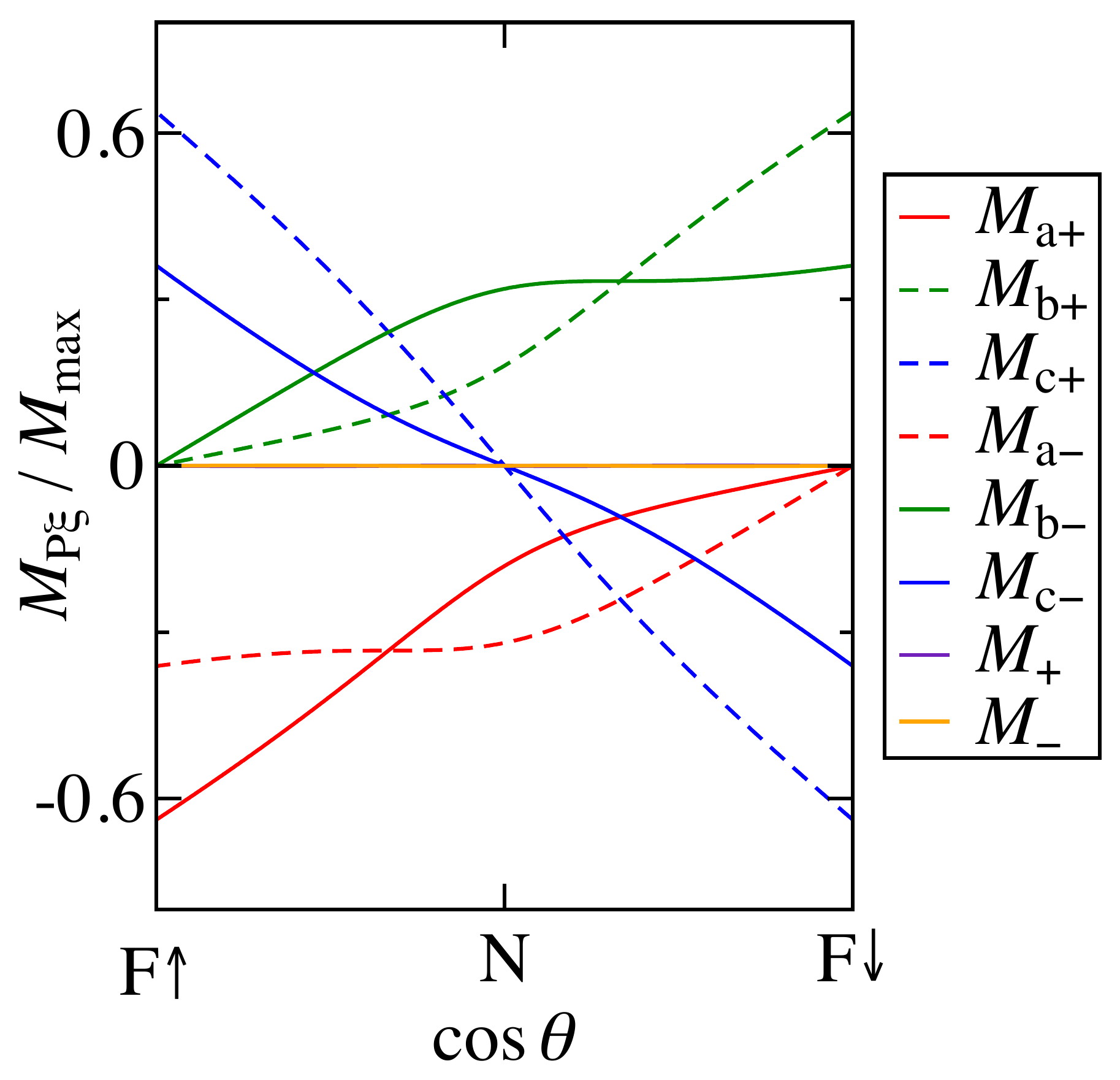}}
  \end{tabular}
  \caption{\label{fig:model3_xi0} Trimer with two orbitals per site and a noncollinear magnetic structure in the adiabatic approximation, Eq.~\eqref{eq:hblocksp}, and with no relativistic spin-orbit interaction ($t = 1$ and $\xi = 0$).
  (a) Eigenenergies, Eq.~\eqref{eq:energies_model3}.
  (b) Atomic orbital magnetic moment, Eq.~\eqref{eq:magmoms_model3l}.
  (c) Orbital magnetic moment arising from the circulating currents, Eq.~\eqref{eq:magmoms_model3p}.
  The magnetic structures are defined by Eq.~\eqref{eq:magdir} and illustrated in Fig.~\ref{fig:trimer}(b).
  The curves are labelled with the states for F$\uparrow$ taken as reference: the color labels $M_{\MR{P}}$, similarly to $M_k$ in Fig.~\ref{fig:model1}(b); solid lines and dashed lines indicate the sign of $M_{\MR{L}}$.
  The following combinations of eigenstates are also plotted: $(\lambda) = (\MR{a}\lambda) + (\MR{b}\lambda) + (\MR{c}\lambda)$, with $\lambda=\pm$.}
\end{figure}

Now that the analytical expressions have been derived, let us explore the physics.
We begin by examining what happens when the RSOI is turned off ($\xi = 0$), with the results gathered in Fig.~\ref{fig:model3_xi0}.
Consider first the ferromagnetic structures.
There are two pairs of degenerate eigenenergies, and another is non-degenerate, see Fig.~\ref{fig:model3_xi0}(a).
They can be characterized by their atomic orbital moments, Fig.~\ref{fig:model3_xi0}(b), and by their translational motion $M_{\MR{P}}$, Fig.~\ref{fig:model3_xi0}(c).
The degenerate eigenstates have strong $\circlearrowleft$ or $\circlearrowright$ orbital character ($M_{\MR{L}} \approx 1$ or $-1$, respectively), and $k = \pm1$ character ($M_{\MR{P}}/M_{\MR{max}} \approx \mp0.5$).
The non-degenerate eigenstates are orbitally mixed ($M_{\MR{L}} = 0$), with $k = 0$ character ($M_{\MR{P}} = 0$).
There is overall no net orbital magnetic moment, as non-degenerate eigenstates have zero orbital moment, and degenerate ones have orbital moments with opposite values, thus cancelling out.
As already seen in the simpler model of Sec.~\ref{sec:model2}, the noncollinear magnetic structures lift the energy degeneracies and modify the orbital moments of each eigenstate, enabling a net orbital moment without the RSOI being present.
To illustrate this, we sum all the contributions corresponding to the $+$ and $-$ bands, which corresponds to placing three electrons in the three upper or lower eigenstates.
There is a net atomic orbital moment, see Fig.~\ref{fig:model3_xi0}(b), with a $C(\theta)$-like angular dependence (Eq.~\eqref{eq:schir}), but no net current, see Fig.~\ref{fig:model3_xi0}(c).

\begin{figure}[t]
  \centering
  \setlength{\tabcolsep}{2pt}
  \begin{tabular}{lll}
    (a) & (b) & (c) \vspace{-1.5em}\\
  \hspace{1.2em}\includegraphics[width=0.3\textwidth]{{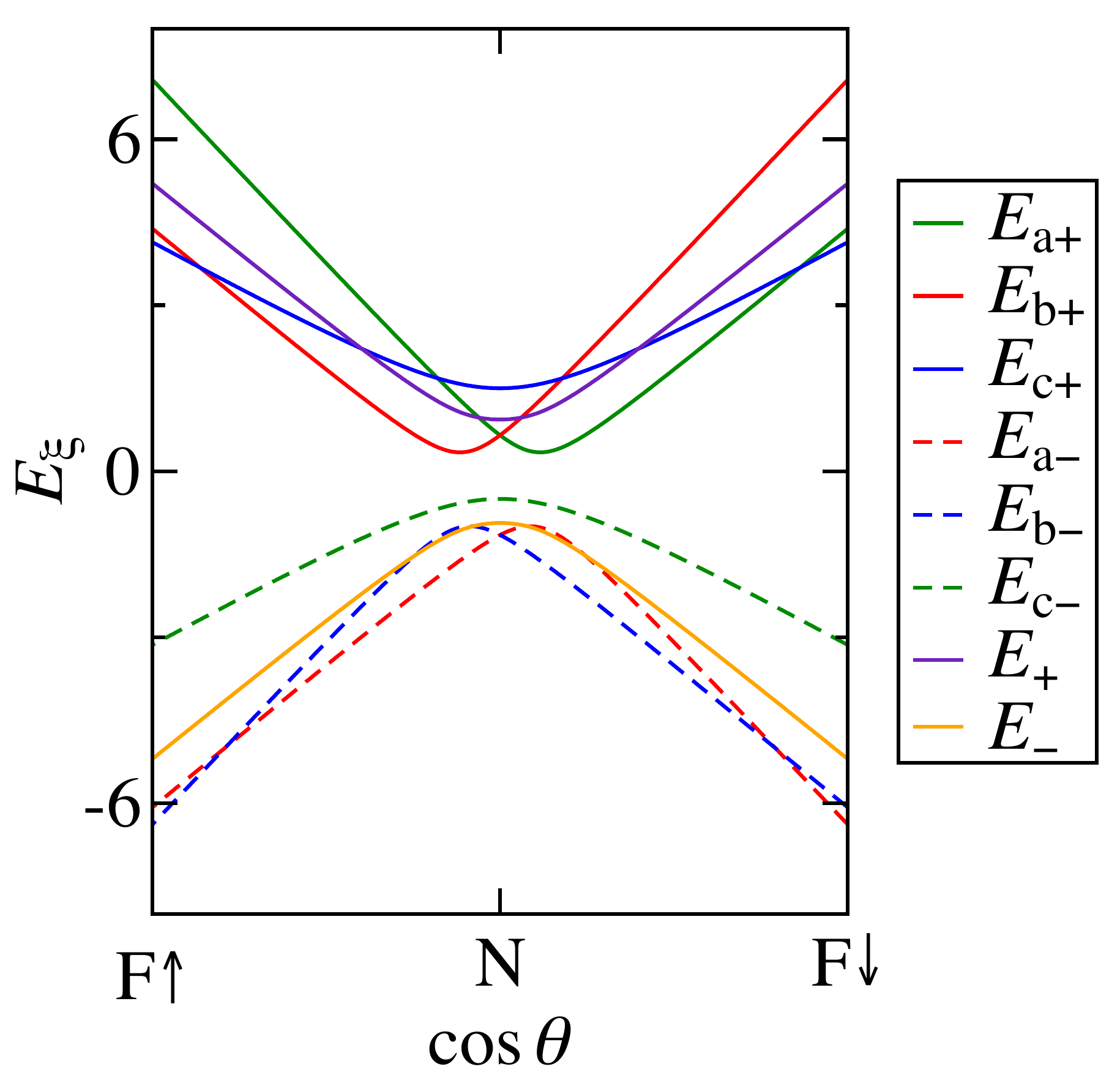}} &
  \hspace{1.3em}\includegraphics[width=0.3\textwidth]{{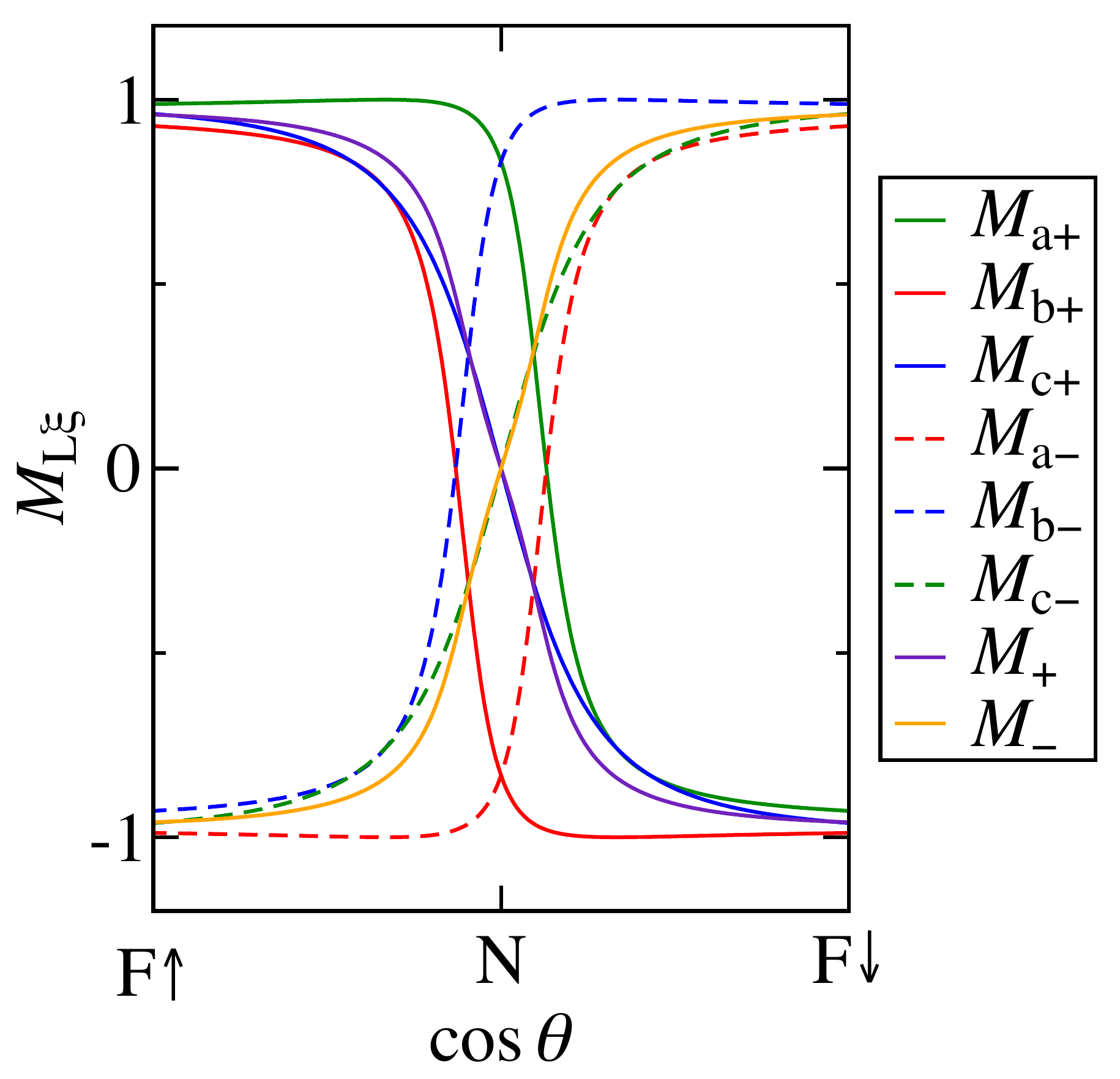}} &
  \hspace{1.2em}\includegraphics[width=0.3\textwidth]{{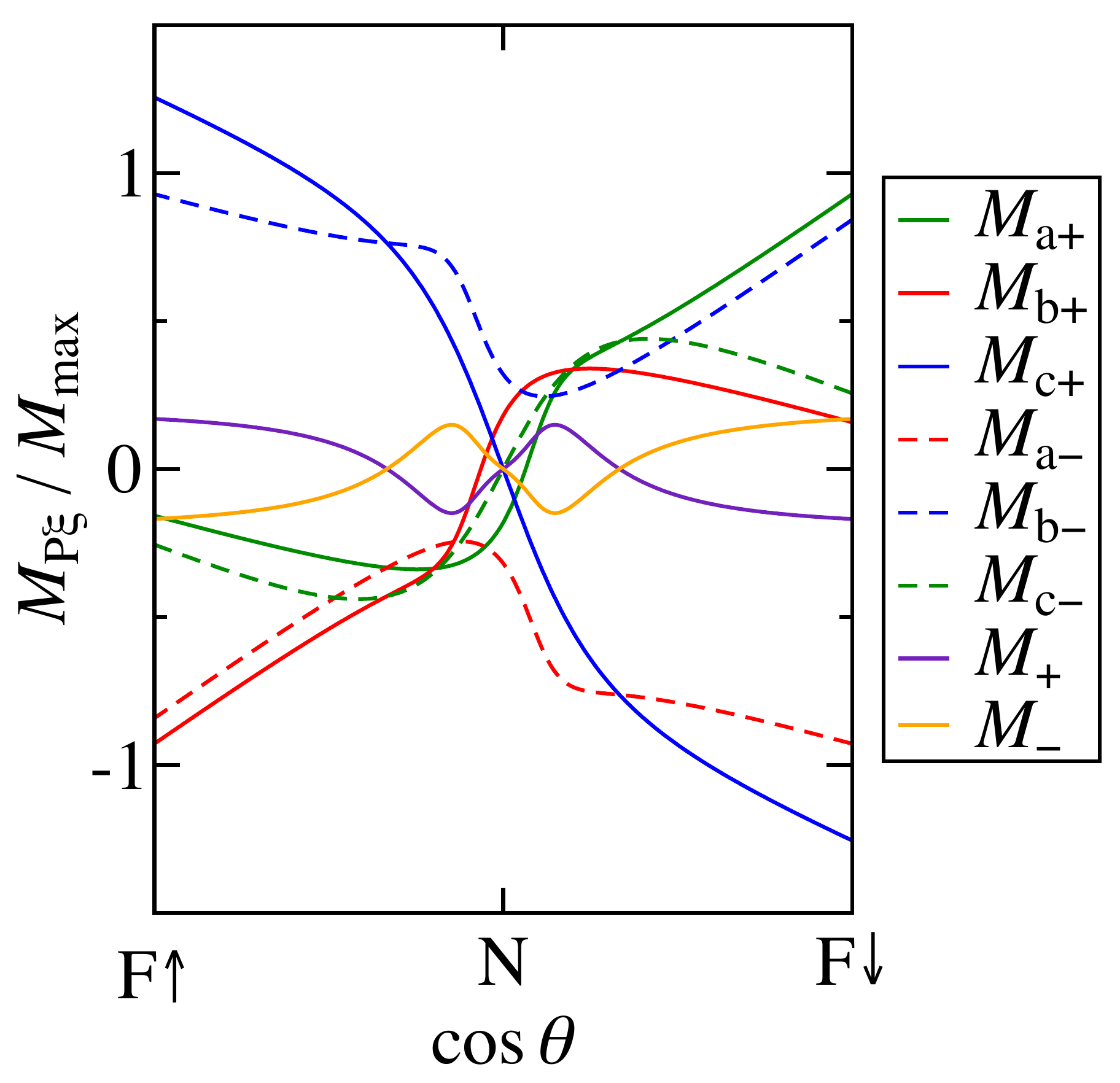}}
  \end{tabular}
  \caption{\label{fig:model3_xi5}  Trimer with two orbitals per site and a noncollinear magnetic structure in the adiabatic approximation, Eq.~\eqref{eq:hblocksp}, and with strong relativistic spin-orbit interaction ($t = 1$ and $\xi = 5$).
  (a) Eigenenergies, Eq.~\eqref{eq:energies_model3}.
  (b) Atomic orbital magnetic moment, Eq.~\eqref{eq:magmoms_model3l}.
  (c) Orbital magnetic moment arising from the circulating currents, Eq.~\eqref{eq:magmoms_model3p}.
  The magnetic structures are defined by Eq.~\eqref{eq:magdir} and illustrated in Fig.~\ref{fig:trimer}(b).
  The curves are labelled with the states for F$\uparrow$ taken as reference: the color labels $M_{\MR{P}}$, similarly to $M_k$ in Fig.~\ref{fig:model1}(b); solid lines and dashed lines indicate the sign of $M_{\MR{L}}$.
  The following combinations of eigenstates are also plotted: $(\lambda) = (\MR{a}\lambda) + (\MR{b}\lambda) + (\MR{c}\lambda)$, with $\lambda=\pm$.}
\end{figure}

We finally bring the RSOI into play.
If it is weak when comparing to the kinetic hopping, $\xi \ll \NORM{t}$, the picture is qualitatively similar to the previous one.
One major difference is that it lifts the energy degeneracies in the ferromagnetic structures, thus allowing net orbital moments.
This is the well-known role of the RSOI in ferromagnetic systems.
We focus on the opposite limit, $\xi \gg \NORM{t}$, to see how it counteracts the kinetic term.
The results for $\xi = 5$ and $t = 1$ are shown in Fig.~\ref{fig:model3_xi5}.
In the adiabatic approximation, the RSOI is projected onto the local magnetization direction.
Combined with our choice of orbitals, this results in a simple $\cos\theta$ dependence, as seen in Eq.~\eqref{eq:hblocksp}.
All the results show the same behavior, except for a small window around the N\'eel magnetic structure, where $\cos\theta = 0$, and the kinetic term becomes important.
The eigenenergies are thus linear in $\cos\theta$, Fig.~\ref{fig:model3_xi5}(a), and the atomic orbital moments are almost saturated to the atomic limit, see Fig.~\ref{fig:model3_xi5}(b).
The $\xi \gg t$ limit also modifies how the electrons move around the trimer, revealed in the behavior of $M_{\MR{P}}$, Fig.~\ref{fig:model3_xi5}(b).
For the ferromagnetic structures we find values close to those of the model without orbital dependence, $M_{\MR{P}} / M_{\MR{max}} \approx 0, \pm1$, compare with Fig.~\ref{fig:model2_J5}(b), and a linear departure from those values when the magnetic structure departs from the ferromagnetic ones.
In this limit the trimer approximately decouples into two separate orbital channels, each behaving as described in Sec.~\ref{sec:model2}.
When the magnetic structure is close to the N\'eel structure, there is some subtle behavior.
To illustrate this, we sum all the contributions corresponding to the $+$ and $-$ bands, which corresponds to placing three electrons in the three upper or lower eigenstates.
The average atomic orbital moment is featureless, see Fig.~\ref{fig:model3_xi5}(b), but the net current changes sign before vanishing at the N structure, see Fig.~\ref{fig:model3_xi5}(c).

\section{Discussion and conclusions}\label{sec:conclusions}
In this paper we discussed a sequence of related models for a trimer, to ascertain how magnetic noncollinearity leads to orbital magnetism, even in the absence of the usual RSOI.
The simplest model was introduced in Sec.~\ref{sec:model1}, and an external magnetic field was used to define the orbital magnetic moment arising from currents circulating around the trimer.
It was augmented with the spin degrees of freedom in Sec.~\ref{sec:model2}, and a family of noncollinear magnetic structures was found to lead to the same kind of orbital moment, even without an external magnetic field.
The model was finally endowed with orbital degrees of freedom in Sec.~\ref{sec:model2}, enabling the appearance of the RSOI.
The adiabatic approximation was adopted, and the competition between the bond-forming tendencies of the orbital-dependent hopping, and the favoring of current-carrying states by the magnetic noncollinearity and the RSOI was analyzed.

Trimer-like structures have been considered in the seminal work of Ref.~\onlinecite{Shindou2001} ($J \gg \NORM{t}$) and of Refs.~\onlinecite{Tatara2002,Tatara2003,Tatara2003a} ($J \ll \NORM{t}$), where the appropriate limits of our model are indicated.
Those works established the scalar spin chirality $C(\theta)$, see Eq.~\ref{eq:schir}, as the smoking gun of the non-RSOI-driven orbital effects.
It vanishes for ferromagnetic structures and for the triangular antiferromagnetic N\'eel structure.
Our results show that the orbital magnetism of an individual eigenstate is not proportional to $C(\theta)$ (for instance, some have a pure $\cos\theta$ dependence), but that considering a full `shell' or `band' does yield this angular dependence, both in the $J \ll \NORM{t}$ and in the $J \gg \NORM{t}$ limits.
We thus expect partial electron fillings to lead to non-$C(\theta)$ angular behavior, as we already found in DFT calculations for magnetic trimers.~\cite{Dias2016}

We also analyzed separately the behavior of the two contributions to the orbital magnetic moment, the atomic one and the one due to circulating (bound) currents.
The former is derived from the atomic orbital Zeeman interaction, while the latter follows from the Peierls phase acquired by the hopping amplitudes.
In general such a separation is also possible, as established by the modern theory of orbital magnetization.~\cite{Souza2008,Thonhauser2011}
They give access to two aspects of the persistent (bound) current flowing around the trimer: whether it swirls locally around each atomic site (the local orbital moment), and whether there is a net current circulating around the trimer (the nonlocal orbital moment).
Our previous work in Ref.~\onlinecite{Dias2016} focused on the atomic orbital moment in trimers but also in a skyrmion lattice, where a topological contribution was identified, and found to be separable from the RSOI-driven one.
As this arose from the magnetic noncollinearity being of a special type for a skyrmion, as encoded in its topological charge,~\cite{Nagaosa2013} we expect that also the nonlocal orbital moment of skyrmions should also contain such a topological contribution.~\cite{Lux2017}
The orbital magnetic moment can be measured independently of the spin magnetic moment with x-ray magnetic circular dichroism,~\cite{Thole1992,Carra1993,Chen1995} and there is a theoretical proposal for how to separate the local and nonlocal contributions to the orbital moment.~\cite{Souza2008}

Recent advances in atomic scale manipulation with the tip of a scanning tunneling microscope have enabled \`a la carte assembly of magnetic nanostructures, including trimers.~\cite{Hirjibehedin2006,Loth2012,Hermenau2017}
The several physical regimes explored in our model can be realized experimentally: the interplay between $J$ and $\NORM{t}$ can be tuned by changing the separating between the magnetic atoms, or by assembling them on metallic or (semi-)insulating surfaces, while the strength of the RSOI, $\xi$, can be manipulated by choosing a surface with strong RSOI, or by working with heavy magnetic atoms.\cite{Zhang2012a}
Detection of the orbital magnetism at the atomic scale remains challenging, but recent progress in very sensitive magnetometers utilizing nitrogen vacancies in nanodiamonds might open a way forward.~\cite{Rondin2014}

For a long time the experimental and theoretical study of the orbital magnetic moment has been neglected in favor of its spin counterpart.
This is natural, as the spin moment in most cases determines most of the total magnetic moment in a solid, and the magnetic structures and dynamics are governed by the interatomic spin exchange interactions.
Orbital interactions are well-known to be important for transport measurements, as can be seen from the large family of Hall effects.
The recent focus on the coupling between the itinerant electrons and the spin moments, described by emergent electromagnetic fields, is part of that.~\cite{Xiao2010a,Nagaosa2013}
We hope that our work helps bringing the humble orbital magnetic moment back to the limelight.
\vspace{5em}

\appendix
\section{Eigenvectors for the two-dimensional problem}\label{app:spindiag}
We wish to diagonalize the following matrix with real parameters $w$, $x$, $y$ and $z$,
\begin{equation}
  A = \begin{pmatrix} a + b_z & b_x - \iu\,b_y \\ b_x + \iu\,b_y & a - b_z \end{pmatrix}
  = a\,\sigma_0 + \VEC{b}\cdot\bm{\upsigma} \quad.
\end{equation}
The eigenvalues and the associated eigenspace projectors are then
\begin{equation}
  \lambda_\pm = a \pm \sqrt{\VEC{b}\cdot\VEC{b}} \quad,\qquad
  P_{\pm} = \frac{1}{2}\left(\sigma_0 \pm \frac{\VEC{b}\cdot\bm{\upsigma}}{\sqrt{\VEC{b}\cdot\VEC{b}}}\right) \quad .
\end{equation}
The corresponding eigenvectors can be parametrized as
\begin{equation}
  \ket{+} = \begin{pmatrix} c \\ e^{\iu\varphi} s \end{pmatrix} \quad,\qquad \ket{-} = \begin{pmatrix} -e^{-\iu\varphi} s \\ c \end{pmatrix} \quad ,
\end{equation}
with
\begin{equation}
  c = \sqrt{\frac{1 + \cos\theta}{2}} \quad,\qquad s = \sqrt{\frac{1 - \cos\theta}{2}} \quad,
\end{equation}
and the angles
\begin{equation}
  \cos\theta = \frac{b_z}{\sqrt{\VEC{b}\cdot\VEC{b}}} \quad,\qquad \theta \in [0,\pi] \quad,\qquad
  \tan\varphi = \frac{b_y}{b_x} \in [0,2\pi] \quad.
\end{equation}

\acknowledgments
We thank Juba Bouaziz, Phivos Mavropoulos, Yuriy Mokrousov and Stefan Bl\"ugel for insightful discussions, and Julen Iba{\~n}ez-Azpiroz and Sascha Brinker for a critical reading of the manuscript.
We would also like to acknowledge the software package Mathematica~\cite{Mathematica} for its assistance with ensuring the correctness of the sometimes cumbersome analytical expressions.
This work is supported by the European Research Council (ERC) under the European UnionÕs Horizon 2020 research and innovation programme (ERC-consolidator grant 681405 -- DYNASORE).

\bibliography{references}

\end{document}